\documentclass[preprint,aps,showpacs,preprintnumbers,amsmath,amssymb,nofootinbib]
{revtex4}
\usepackage{graphicx}
\usepackage{epsfig}
\begin{document}

\begin{flushright}
%\preprint{hep-ph/xxyyzz}
%\today\\
%UH-0528
\end{flushright}

%%%%%%%%%%%%%%%%%%%%%%%%%%%%%%%%%%%%%%%%%%%%%%%%%%%%%%%%%%%%%%%%%%%%%%%%

\newcommand{\be}{\begin{equation}}
\newcommand{\ee}{\end{equation}}
\newcommand{\bea}{\begin{eqnarray}}
\newcommand{\eea}{\end{eqnarray}}
\newcommand{\nn}{\nonumber}
\def\CP{{\it CP}~}
\def\cp{{\it CP}}
\title{\large Predicting Leptonic CP phase by considering deviations in charged lepton and neutrino sectors}
%\title{\large Constraining the CP Dirac violating phase  with different neutrino mixing patterns}
\author{Sruthilaya M.,  Soumya C., K. N. Deepthi, R. Mohanta }
\affiliation{
School of Physics, University of Hyderabad, Hyderabad - 500 046, India }

%\date{\today}

\begin{abstract}
Recently, the reactor mixing angle $\theta_{13}$  has been  measured precisely by Daya Bay, RENO
and T2K experiments with a  moderately large value. However, the standard form of neutrino mixing patterns such as bimaximal, 
tri-bimaximal, golden ratio of types A and B, hexagonal  etc., 
which are based on certain flavor symmetries, predict vanishing  $\theta_{13}$.
Using the fact that the neutrino
mixing matrix can be represented as  $V_{\rm PMNS}=U_l^{\dagger} U_\nu P_\nu$, where $U_l$ and $U_\nu$ result from the diagonalization
of the charged lepton and  neutrino mass matrices and $P_\nu$ is a diagonal matrix containing Majorana phases, we explore the possibility
of accounting for the large reactor mixing angle  by considering deviations both in the charged lepton and neutrino sector.
In the charged lepton sector we consider the deviation as an additional rotation in the (12) and (13) planes,
 whereas in neutrino sector we consider deviations to various neutrino mixing patterns through
(13) and (23) rotations. We find that with the  inclusion of these deviations it is possible to accommodate the observed  large 
reactor mixing angle $\theta_{13}$, and one can also 
 obtain limits on the CP violating Dirac phase $\delta_{CP}$ and Jarlskog invariant $J_{CP}$ for most of the cases. We then explore whether 
our findings can be tested in the currently running NO$\nu$A experiment with 3 years of data taking in neutrino mode  followed by 3 years 
with anti-neutrino mode.
\end{abstract}

\maketitle

\section{Introduction}

The phenomenon of neutrino oscillation is found to be the first substantial evidence for physics beyond standard model.
The results from various neutrino oscillation experiments \cite{osc}, established the fact that the three flavors of neutrinos mix with each other as
they propagate and form the mass eigenstates. The mixing is described by Pontecorvo-Maki-Nakagawa-Sakata (PMNS) matrix $V_{PMNS}$ \cite{pmns}
analogous to the CKM mixing matrix in the quark sector. The mixing matrix is unitary, and hence,   parameterized in terms of three rotation angles
$\theta_{12}^{}$, $\theta_{23}^{}$, $\theta_{13}^{}$ and three CP-violating phases, one Dirac type ($\delta_{CP}$) and two Majorana types ($\rho, \sigma$) as
\be
V_{PMNS} \equiv U_{PMNS}.P_\nu = \left( \begin{array}{ccc} c^{}_{12} c^{}_{13} & s^{}_{12}
c^{}_{13} & s^{}_{13} e^{-i\delta_{CP}} \\ -s^{}_{12} c^{}_{23} -
c^{}_{12} s^{}_{13} s^{}_{23} e^{i\delta_{CP}} & c^{}_{12} c^{}_{23} -
s^{}_{12} s^{}_{13} s^{}_{23} e^{i\delta_{CP}} & c^{}_{13} s^{}_{23} \\
s^{}_{12} s^{}_{23} - c^{}_{12} s^{}_{13} c^{}_{23} e^{i\delta_{CP}} &
-c^{}_{12} s^{}_{23} - s^{}_{12} s^{}_{13} c^{}_{23} e^{i\delta_{CP}} &
c^{}_{13} c^{}_{23} \end{array} \right) P^{}_\nu \;
,\label{standpara}
%     (1)
\ee
where $c^{}_{ij}\equiv \cos \theta^{}_{ij}$, $s^{}_{ij} \equiv \sin
\theta^{}_{ij}$ and $P_\nu^{} \equiv \{ e^{i\rho}, e^{i\sigma}, 1\}$ is
a diagonal phase matrix, which is physically relevant if neutrinos
are Majorana particles.

The solar and atmospheric neutrino  oscillation parameters are precisely known from various neutrino oscillation experiments.
Recently the reactor mixing angle has also been measured by the Double Chooz \cite{double}, 
Daya Bay \cite{daya-bay, daya-bay1}, RENO \cite{reno}, and T2K
\cite{t2k,t2k-result} experiments with a moderately large value. 
After the discovery of sizable  $\theta_{13}$, much attention has been paid to determine the CP-violation effect in the
lepton sector, in the currently running as well as in   future  long-baseline neutrino oscillation experiments.
As $\theta_{13}$ is non-zero, there could be CP-violation in the lepton sector,
analogous to the quark sector, provided the CP violating phase $\delta_{CP}$ is not vanishingly
small. Hence, it is of  particular importance to determine the  Dirac CP phase $\delta_{CP}$ both theoretically
and experimentally.  
The global analysis of  various neutrino oscillation data has been performed by 
various groups \cite{gfit1,gfit2,gfit3,gfit3a}, and the hint for non-zero $\delta_{CP}$ was anticipated in Refs. \cite{gfit3,gfit3a}. 
Including the data from T2K and Daya Bay, Forero {\it et al.} \cite{gfit4} performed a global fit  and found a hint for non-zero value of 
$\delta_{CP}$ and a deviation of $\theta_{23}$ from $\pi/4$, with the best fit values as $\delta_{CP} \simeq 3 \pi/2$ and $\sin^2 \theta_{23}\simeq 0.57$.
The best fit values along with their 3$\sigma$ ranges of various oscillation parameters from Ref. \cite{gfit4}, are presented in Table-1.
\begin{table}[htb]
\begin{center}
\vspace*{0.1 true in}
\begin{tabular}{|c|c|c|}
\hline
 Mixing Parameters & Best Fit value & $ 3 \sigma $ Range  \\
\hline
$\sin^2 \theta_{12} $ &~ $0.323$ ~& ~$ 0.278 \to 0.375 $~\\

$\sin^2 \theta_{23}  $ (NO) &~ $0.567$ ~& ~$ 0.392 \to 0.643 $~\\

$\sin^2 \theta_{23}  $ (IO) &~ $0.573$ ~& ~$ 0.403 \to 0.640 $~\\

$\sin^2 \theta_{13} $ (NO) &~ $0.0234$ ~& ~$ 0.0177 \to 0.0294 $~\\

$\sin^2 \theta_{13} $ (IO) &~ $0.0240$ ~& ~$ 0.0183 \to 0.0297 $~\\

$\delta_{\rm CP}$ (NO) & ~$1.34 \pi$ & $ ~(0 \to 2 \pi)~ $\\
$\delta_{\rm CP}$ (IO) & ~$1.48 \pi$ & $ ~(0 \to 2 \pi)~ $\\
$\Delta m_{21}^2/ 10^{-5} ~{\rm eV}^2 $ & $ 7.60 $ & $ 7.11 \to 8.18 $ \\

$\Delta m_{31}^2/ 10^{-3}~ {\rm eV}^2 ~({\rm NO}) $~ &~ $ 2.48
$ & $ 2.3 \to 2.65 $ \\

$\Delta m_{32}^2/ 10^{-3}~ {\rm eV}^2~ ({\rm IO}) $ ~&~ $ -2.38 $ & $ -2.54 \to -2.20 $ \\

\hline
\end{tabular}
\end{center}
\caption{The best-fit values and the  $3\sigma$ ranges of the neutrino oscillation parameters from Ref.  \cite{gfit4}.}
\end{table}

Understanding the origin of the patterns of neutrino masses and mixing, emerging from the neutrino oscillation data is one of the
most challenging problems in neutrino physics. In fact, it is part of the more fundamental problem of particle physics of
 understanding the origin of masses and the mixing pattern in quark and lepton sector.
As we know, the phenomenon of neutrino oscillation is characterized by two
large mixing angles, the solar ($\theta_{12}$) and the atmospheric ($ \theta_{23}$), and one not
so large reactor mixing angle $\theta_{13}$. Initially it was believed  that the reactor mixing angle would  be
vanishingly small and motivated by such anticipation many models were proposed to explain the neutrino mixing
pattern which are generally based on some kind of discrete flavor symmetries like $S_3$, $S_4$, $A_4$, etc. \cite{discrete, discrete1,discrete2}.
For an example, the tri-bimaximal (TBM) mixing  pattern \cite{tbm} is one such well motivated model having $\sin^2 \theta_{12}=\frac{1}{3}$ and
$\sin^2 \theta_{23}=\frac{1}{2}$, which plays a crucial role for model building.
However, in TBM mixing pattern  the value of $\theta_{13}$ is zero and the CP phase $\delta_{CP}$ is consequently undefined.
After the experimental discovery  of moderately large  $\theta_{13}$, various perturbation  terms are added to
the TBM mixing pattern and it was found that it can still be used  to describe the neutrino mixing pattern or
model building with suitable modifications \cite{xing}.

Thus, in a nutshell the experimental discovery of moderately large value of  reactor mixing angle  caused a profound change in the subject
of flavor models,  describing leptonic mixing.  Some of the models are outdated while  others are suitably modified by
including appropriate perturbations/corrections to accommodate the observed value of  $\theta_{13}$ \cite{th13}. In this paper,
 we would like to consider the effect
of perturbations to few such well motivated models   which are based on
certain discrete flavour symmetries like $A_{4}$, $\mu - \tau$, etc. These models
include tri-bimaximal mixing (TBM) \cite{tbm}, bi-maximal mixing (BM) \cite{bm}, golden ratio type A (GRA) \cite{gra,gra1}, golden 
ratio type B (GRB) \cite{GRA/B, petcov}, 
hexagonal (HG) \cite{HG} mixing patterns.
However, as we know these forms do not accommodate non-zero value for the  reactor  mixing angle $\theta_{13}$,
and hence need to be modified suitably to provide the leptonic mixing angles  in compatible with the experimental data.
In this paper, we are interested to look for such a possibility. Although, this aspect has been widely
studied  in the literature, see for example \cite{th13,xing,petcov,petcov1,charged-lepton, charged-lepton1}, the main difference between the previous studies
and our work is, we have considered a very simple form of deviation matrix in terms of minimal number
of new independent parameters, which can provide corrections to both charged lepton and the standard
neutrino mixing matrices. This in turn not only accommodates the observed mixing angles but also constrains the Dirac CP violating phase
$\delta_{CP}$.

It is well-known that the determination of  the CP violating phases and in particular
the Dirac CP phase $\delta_{CP}$ is an  important issue in the study of neutrino physics. 
Many dedicated long-baseline experiments are planned to study CP violation in the neutrino sector.
The theoretical prediction for the determination of CP phase in the neutrino mixing matrix depends on the approach as well as the type of
symmetries one uses to understand the pattern of neutrino mixing.
Obviously a sufficiently precise measurement of $\delta_{CP}$ will serve as a very useful constraint  for identifying the approaches
and symmetries, if any. In this work, we would also like to explore
whether it is possible to constrain the CP phase $\delta_{CP}$ by considering corrections 
to the leading order charged lepton and neutrino mixing matrices and if so whether it is possible to verify such predictions with the
data from ongoing NO$\nu$A experiment.

The paper is  organized as follows. In section II we  present the basic framework of our analysis. The deviations to the various mixing patterns due to
neutrino and charged lepton sectors are discussed in Sections III and IV respectively. Section V contains summary and conclusion.

\section{Framework}

It is well-known that the lepton mixing matrix arises from the overlapping of the matrices that diagonalize
charged lepton and neutrino mass matrices i.e.,
\bea
U_{\rm PMNS} = U_l^\dagger U_\nu \label{mix}\;.\eea
For the study of leptonic mixing  it is generally assumed that the charged lepton mass matrix is diagonal and hence, the corresponding
mixing matrix $U_l$ be an identity
matrix. However, the neutrino mixing matrix $U_\nu$ has a specific form dictated by the symmetry
which generally  fixes the values of the three mixing angles in $U_\nu$. The small deviations of
the predicted values of the mixing angles from their corresponding  measured values  are considered, in general, as  perturbative corrections
arising from symmetry breaking effects.  A variety  of symmetry forms of $U_\nu$ have been explored in the literature e.g.,
tri-bimaximal (TBM), bi-maximal (BM), Golden ratio type-A (GRA), type-B (GRB), hexagonal (HG) and so on.
All these mixing patterns can be written in a generalized form as shown in Ref. \cite{petcov}.
For the case of TBM, BM, GRA, GRB and HG forms for $U_\nu$, one can have $\theta_{23}^\nu =-\pi/4$, while
$\theta_{12}^\nu$ takes the values $\sin^{-1}(1/\sqrt 3 )$, $\pi/4$, $\sin^{-1}(1/\sqrt{2+r})$, $\sin^{-1}(\sqrt{3-r}/2)$ ($r$ being the golden
ratio i.e., $r=(1+\sqrt{5})/2$) and $\pi/6$ respectively.
Thus, the generalized neutrino matrix $U_\nu$ corresponding to these cases has
the  form \cite{petcov} 
\be
U_\nu^0=
\left ( \begin{array}{ccc}
\cos \theta_{12}^\nu      & \sin  \theta_{12}^\nu    & 0 \\
-\frac{\sin \theta_{12}^\nu}{\sqrt{2}}      & \frac{\cos  \theta_{12}^\nu}{\sqrt{2}}    & -\frac{1}{\sqrt{2}} \\
-\frac{\sin \theta_{12}^\nu}{\sqrt{2}}      & \frac{\cos  \theta_{12}^\nu}{\sqrt{2}}    & \frac{1}{\sqrt{2}} \\
\end{array}
\right ).\label{gen}
\ee
The superscript `0' is introduced to label the mixing matrix as the leading order matrix arising from certain
discrete flavor symmetries.
A common feature of these mixing matrices is  that, they predict $\theta_{23}= \pm \pi/4$ and $\theta_{13}=0$, if the charged
lepton mixing matrix is considered to be a $3 \times 3 $ identity matrix. However, they differ in their prediction for the
solar mixing angle $\theta_{12}$, which has the value as $\sin^2 \theta_{12}=0.5 $ for BM form, 
$\sin^2 \theta_{12}=1/3$ for TBM, $\sin^2 \theta_{12}=0.276 $ and 0.345
for GRA and GRB mixing and $\sin^2 \theta_{12}=0.25$ for HG mixing patterns.
Thus, one possible way to generate corrections for the mixing angles such that all the mixing angles
$\theta_{23}$, $\theta_{12}$ and $\theta_{13} $  should be  compatible with the observed experimental data, 
 is to include suitable perturbative corrections to both the charged lepton and neutrino mixing matrices $U_l$ and $U_\nu$
respectively. In this paper we are interested to explore such a possibility. 
While considering the corrections to the neutrino
mixing matrix, we assume the charged lepton mixing matrix to be identity matrix and for correction to the charged lepton mass matrix we consider
the neutrino mixing matrix to be either of   TBM/BM/GRA/GRB/HG forms.
Furthermore, we will neglect possible corrections to $U_{\nu}$ from higher dimensional operators and from renormalization group effects.

\section{Deviation in Neutrino sector}
In this section, we consider the corrections to the neutrino mixing matrix such that it can be written as
\be
U_\nu= U_\nu^0  U_{\nu}^{corr}\;,
\ee
where $U_\nu^0$ is one of the  symmetry forms of the mixing matrix as described in Eq. (\ref{gen}) and $U_{\nu}^{corr}$
is a unitary matrix describing the correction to $U_\nu^0$. 
An important requirement is that the correction due to the matrix $U_\nu^{corr}$ should allow sizable deviation of the angle
$\theta_{13}$ from zero and also the required deviations to $\theta_{23}$ and $\theta_{12}$, so that
all the mixing angles should be compatible with their measured values.
As discussed in Ref. \cite{ref111}, $U_{\nu}^{corr}$ can be expressed as $V_{23}V_{13}V_{12}$, where $V_{ij}$ are the rotation
matrices in $(ij)$ plane and hence, can be parameterized by three mixing angles and one phase. In this work, we consider
the simplest case of such perturbation which involves only minimal set of new independent parameters, i.e., we consider the deviations involving only 
two new parameters (one rotation angle and one phase), which basically corresponds to  perturbation induced by a single rotation.
There are several variants of this approach exist in the literature, generally for TBM mixing pattern \cite{xing}. 
The main difference between the previous studies and our work is that apart from predicting the values of the
mixing angles compatible with their experimental range,  we have also looked into the
possibility of constraining the CP phase $\delta_{CP}$, not only for TBM case, but also for other
variety of mixing patterns.

\subsection{Deviation due to 23 rotation}
First, we would like to consider additional rotation in the 23 plane. Since the charged lepton mixing matrix is considered to be identity
in this case, 
the PMNS mixing matrix can be obtained by
multiplying the neutrino mixing matrix $U_{\nu}^0$ with the 23 rotation matrix as follows
\be
U_{PMNS} = U_{\nu}^0  \left ( \begin{array}{ccc} 1     & 0    & 0 \\
0    & \cos \phi    & e^{-i  \alpha} \sin \phi \\
0   & -e^{i \alpha} \sin \phi     & \cos \phi\\
\end{array}
\right )\;,\label{type1}
\ee
where $\phi$ and $\alpha$ are arbitrary free parameters.  The mixing angles  $\sin^2 \theta_{12}$, $\sin^2 \theta_{23}$ and $\sin \theta_{13}$ can be obtained
using the relations
\begin{eqnarray}
  \sin^{2}\theta_{12}=\frac{|U_{e2}|^{2}}{1-|U_{e3}|^{2}}~,\qquad \qquad
   \sin^{2}\theta_{23}=\frac{|U_{\mu3}|^{2}}{1-|U_{e3}|^{2}}~,\qquad \qquad
  \sin\theta_{13}=|U_{e3}|~.
 \label{mixing1}
 \end{eqnarray}
Using Eqs. (\ref{gen}), (\ref{type1}) and (\ref{mixing1}), one obtains the mixing angles as
\begin{eqnarray}
 \sin \theta_{13} &=  &\sin \theta_{12}^\nu \sin \phi\;,\label{res-a} \\
\sin^2 \theta_{12} & = & \frac{ \sin^2 \theta_{12}^\nu  \cos^2 \phi}{1-\sin^2 \theta_{12}^\nu \sin^2 \phi}\;,\label{res-b}\\
\sin^2 \theta_{23} &=& \frac{1}{2} \left [\frac{1- \sin^2 \theta_{12}^\nu \sin^2 \phi - \cos \theta_{12}^\nu \sin 2 \phi \cos \alpha}{1-
\sin^2 \theta_{12}^\nu \sin^2 \phi}\right ]\;.\label{res1}
\end{eqnarray}
Thus, from Eqs. (\ref{res-a}-\ref{res1}), one can see that by including the 23 rotation matrix as a perturbation, it is possible to have nonzero $\theta_{13}$,
deviation of $\sin^2 \theta_{23}$ from 1/2 and  $\sin^2 \theta_{12}$ from $\sin^2 \theta_{12}^\nu$.
With Eqns. (\ref{res-a}) and (\ref{res-b}) one can obtain the relation between $\sin^2 \theta_{12}$ and $\sin^2 \theta_{13}$ as
\be
\sin^2 \theta_{12} = \frac{\sin^2 \theta_{12}^\nu - \sin^2 \theta_{13}}{1-\sin^2 \theta_{13}}\;.\label{res-ca}
\ee
Thus, it can be seen that in this case one can have $\sin^2 \theta_{12} < \sin^2 \theta_{12}^\nu $, although the deviation is not significant.
Therefore, the BM, GRA and HG forms of neutrino mixing patterns cannot accommodate the observed value of $\sin^2 \theta_{12}$ within its 
$3\sigma$ range.

Furthermore, as we have a non-vanishing and largish $\theta_{13}$, this in turn implies that it could in principle be possible to observe
CP violation in the lepton sector analogous to the quark sector, provided the CP violating phase is not vanishingly
small, in the  long-baseline neutrino oscillation experiments. The Jarlskog invariant,
which is a measure of CP violation,  has the expression in the standard parameterization as
 \begin{eqnarray}
  J_{\rm CP}\equiv{\rm Im}[U_{e 1}U_{\mu 2}U^{\ast}_{\mu 1}U^{\ast}_{e 2 }]=
\frac{1}{8}\sin 2 \theta_{12} \sin 2 \theta_{23} \sin 2 \theta_{13} \cos \theta_{13} \sin \delta_{CP}\;,
 \label{JCP}
 \end{eqnarray}
and is sensitive to the Dirac CP violating phase.
With Eq. (\ref{type1}), one can obtain the value of Jarlskog invariant as
\bea
J_{\rm CP} = -\frac{1}{4}\cos  \theta_{12}^\nu \sin^2 \theta_{12}^\nu   \sin 2 \phi \sin \alpha\;.\label{cpv1}
\eea
Thus, comparing the two Eqs. (\ref{JCP}) and (\ref{cpv1}), one can obtain   the expression for $\delta_{CP}$ as
\be
\sin \delta_{CP}= - \frac{(1-\sin^2 \theta_{12}^\nu  \sin^2 \phi) \sin \alpha}{\left[(1-\sin^2  \theta_{12}^\nu  \sin^2 \phi)^2 -
\cos^2 \theta_{12}^\nu \sin^2 2 \phi \cos^2 \alpha \right ]^{1/2}}\;.\label{del-cp}
\ee
For numerical evaluation  we constrain the parameter $\phi$ from the measured value of $\sin \theta_{13}$  and  vary the phase parameter  $\alpha$ within its allowed range, i.e.,
$-\pi \leq \alpha \leq  \pi$. 
With Eq. (\ref{res-a}) and using the $3\sigma$ range of $\sin^2 \theta_{13}$ and the specified value of $\sin^2 \theta_{12}^\nu$,
 we obtain the allowed range of $\phi$ for various mixing patterns as:
 $(10.9 - 14.2)^\circ $ for BM, $(13.3 - 17.5)^\circ $ for TBM, $(14.8 - 19.2)^\circ $ for GRA, $(13.2 - 17.2)^\circ $ for GRB,
and $(15.6 - 20.3)^\circ $ for HG pattern. 
With these input parameters, we present our results in Figure-1.
The correlation plot between $\sin^2 \theta_{12}$ and $\sin^2 \theta_{13}$ is shown in the top  panel where the magenta, red, green, orange and blue  
plots correspond to BM, TBM, GRA, GRB and HG
 mixing patterns respectively. The horizontal and vertical dashed black lines correspond to the best fit values for $\sin^2 \theta_{12}$ and $
\sin^2 \theta_{13}$, whereas the vertical dashed magenta lines represent the $3\sigma$ allowed range of $\sin^2 \theta_{12}$ and the horizontal dot-dashed 
lines correspond
to the same for $\sin^2 \theta_{13}$. As discussed before,  one can see from the figure that, the predicted values of the mixing angles $\sin^2\theta_{12}$ and
$\sin^2\theta_{13}$ lie within their $3 \sigma$ ranges only for TBM and GRB mixing patterns whereas the predicted value of $\sin^2\theta_{12}$ lies outside
its $3 \sigma$ range for BM, GRA and HG mixing patterns. 
With Eq. (\ref{del-cp}), we obtain
the  constraint on $\delta_{CP}$  as shown in the  middle  panel of Figure-1 for 
TBM case, where we have used the $3\sigma$ allowed range of the mixing angles
$\theta_{12}$, $\theta_{23}$ and $\theta_{13}$.   Using the predicted value of $\delta_{CP}$, correlation between the Jarlskog invariant and 
$\sin^2 \theta_{13}/\sin^2 \theta_{23} $
are shown in the bottom panel of Figure-1 for TBM case. The corresponding results  for GRB mixing pattern are almost same as TBM case and hence, are 
 not shown explicitly in the figures. However, the  allowed ranges  of $\delta_{CP}$ and $J_{CP}$ are listed in Table-2. Since BM, GRA and HG mixing patterns 
cannot accommodate the observed mixing angles
as discussed earlier  in this section, the corresponding results are not listed. 
\begin{figure}[!htb]
\includegraphics[width=7.5cm,height=6cm]{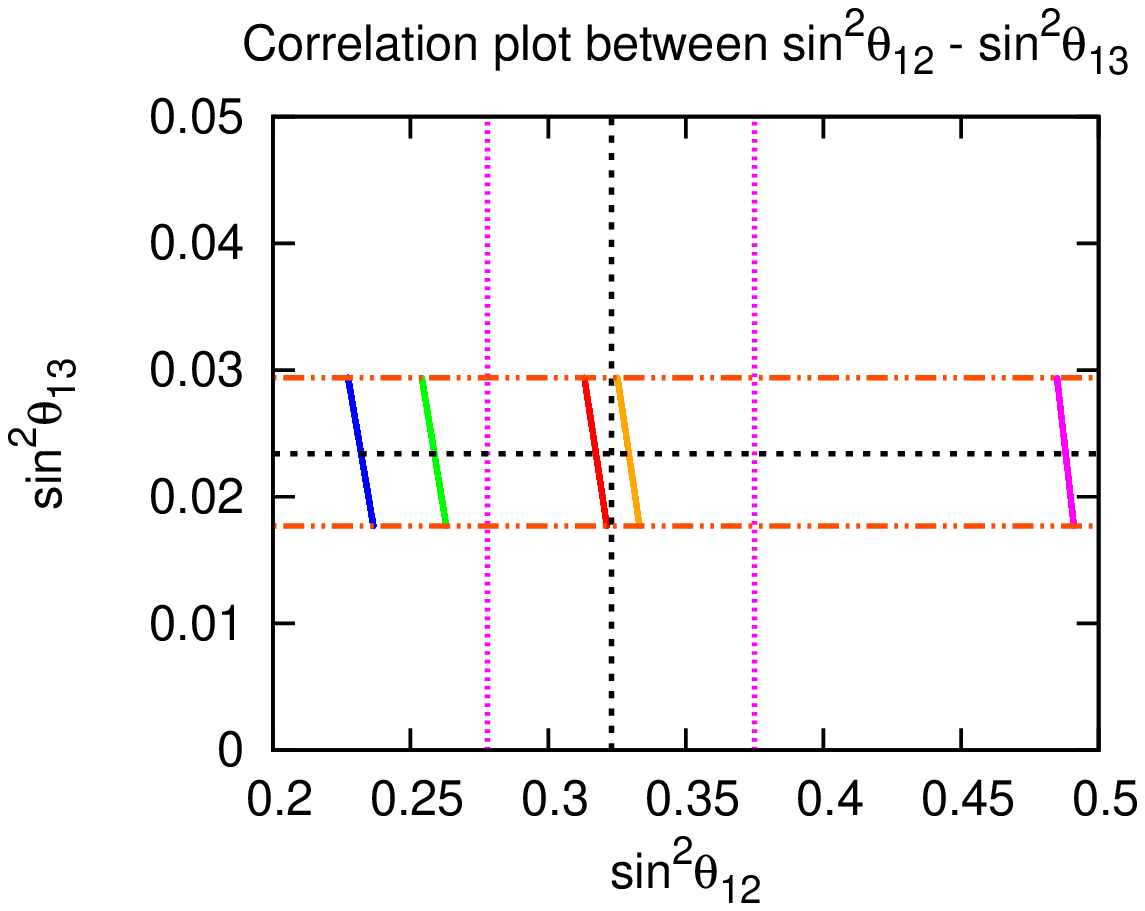}\\
\includegraphics[width=7cm,height=6cm]{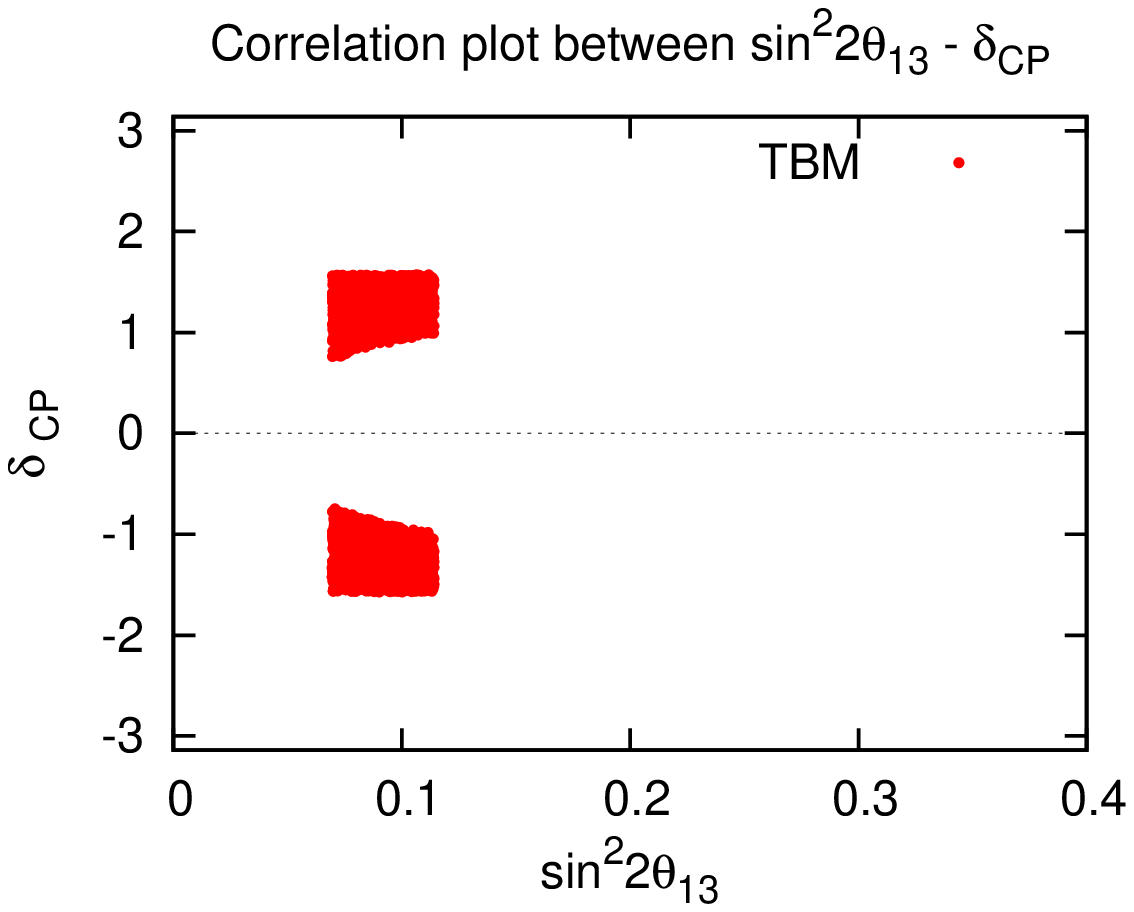}
\includegraphics[width=7.0cm,height=6cm]{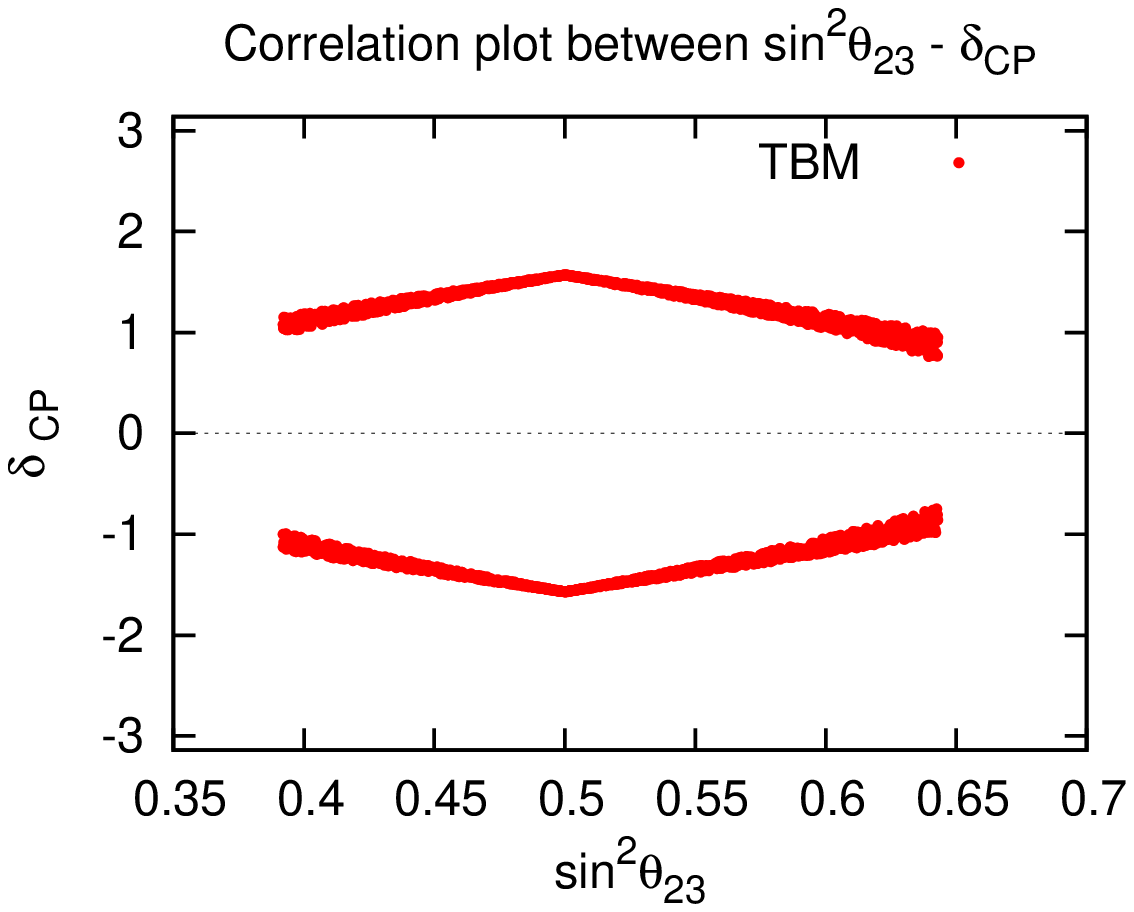}
\includegraphics[width=7.25cm,height=6cm]{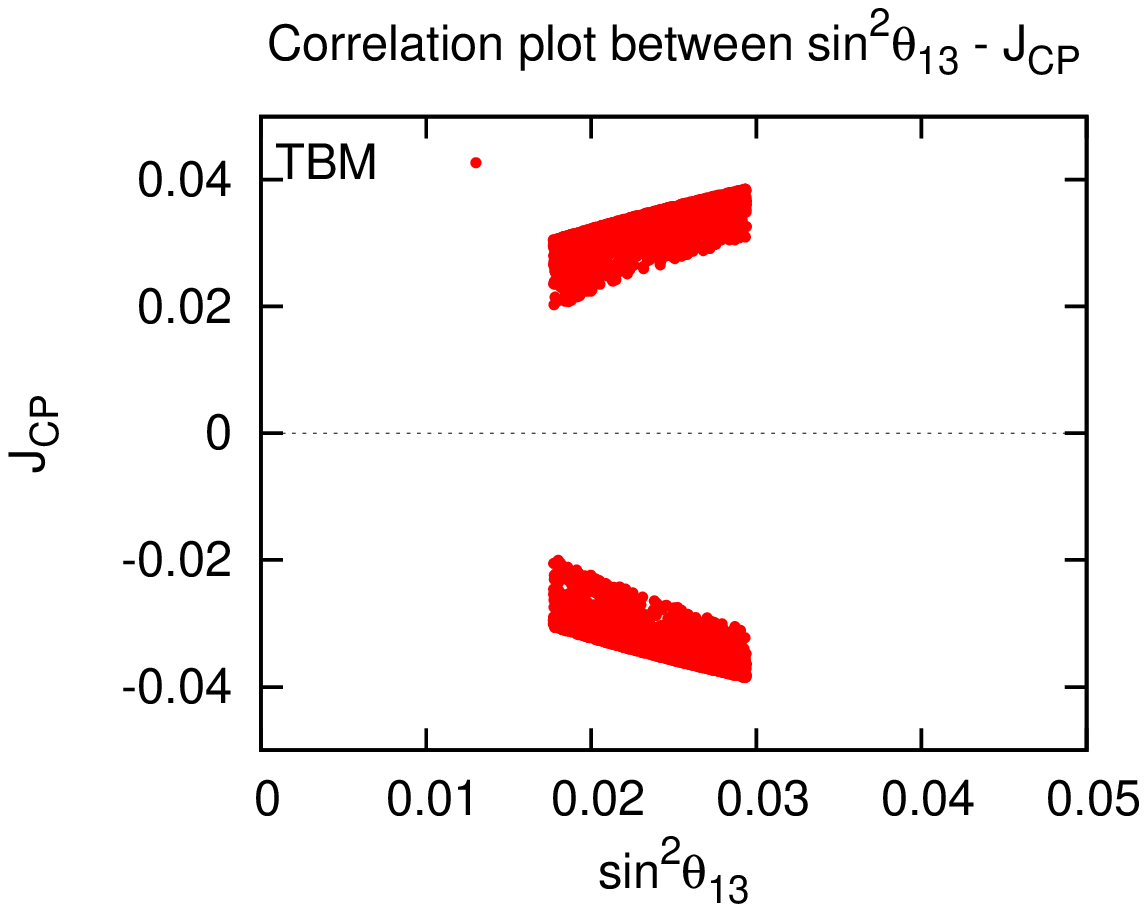}
\includegraphics[width=7.25cm,height=6cm]{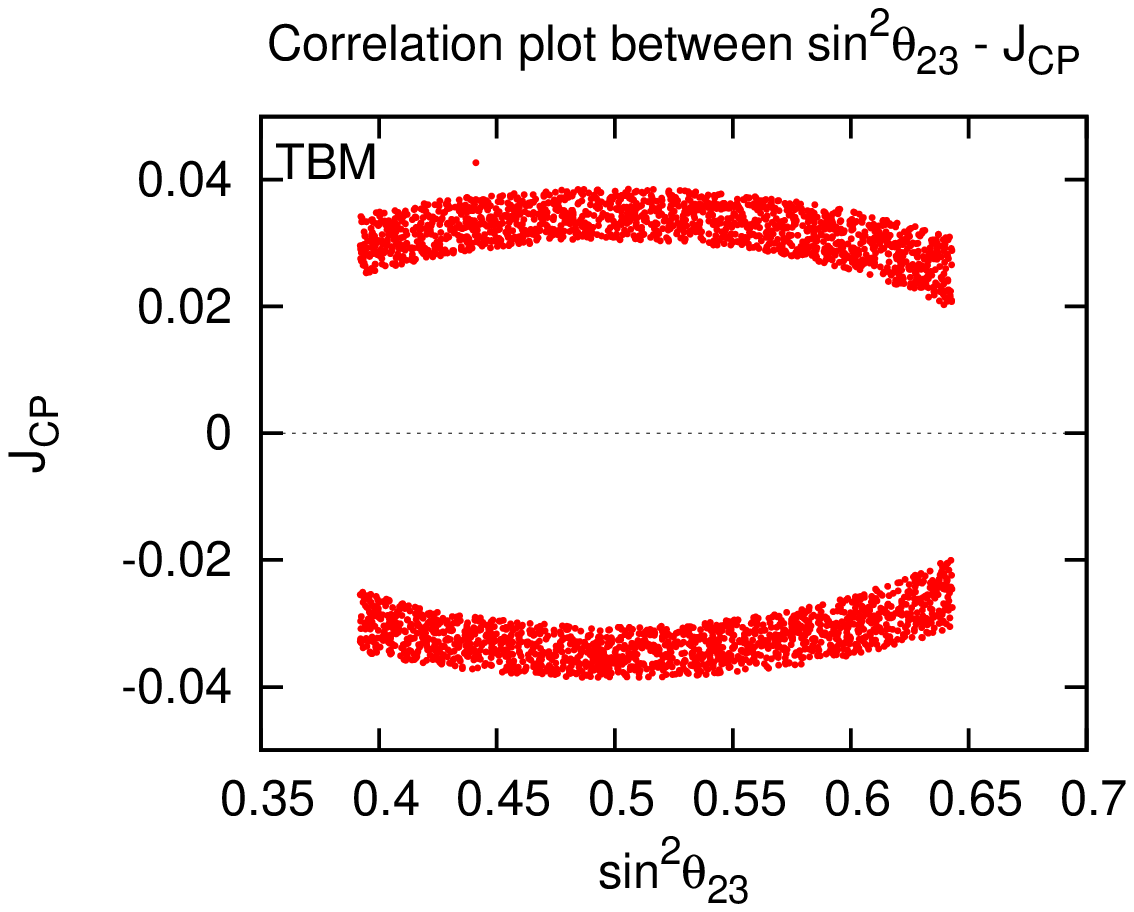}
\caption{Correlation plots between $\sin^2 \theta_{12}$ and $\sin^2 \theta_{13}$ (top  panel) for BM (magenta), TBM (red), GRA (green), GRB (orange) 
and HG (blue) regions.
The horizontal and vertical central lines represent the best fit values where as the dot-dashed orange and dashed magenta 
lines represent corresponding $3 \sigma$
allowed ranges. The constraints on $\delta_{CP}$  for TBM mixing pattern are shown in the  middle  panels and on 
$J_{CP}$ in the bottom panels.}
\end{figure}

Our next objective is to speculate the  possible experimental indications which could support or rule out our findings.
As we know neutrino physics has now entered the precision era  as far as the measured parameters are concerned.
The currently running experiments T2K and NO$\nu$A play a major role in this aspect. These experiments  will provide the precise measurement of atmospheric
neutrino mass square difference and the mixing angle $\theta_{23}$ through
$\nu_\mu$ disappearance channel. They also intend to measure $\theta_{13}$, the CP violation phase $\delta_{CP}$ through
$\nu_\mu$ to $\nu_e$ appearance. Furthermore, NO$\nu$A can potentially resolve the mass-ordering through matter effects
as it has a long-baseline. In this work, we would like to see whether the constraints obtained on $\delta_{CP}$
in our analysis could be probed in the NO$\nu$A experiment with  3 years of data taking with neutrino mode and then followed by another
3 years with antineutrino mode. For our study we do the simulations   using GLoBES \cite{1,2}.

\subsection{Simulation details}

 NO$\nu$A (NuMI Off-axis $\nu_e$ Appearance) is an off-axis long-baseline experiment \cite{nova,nova1}, which uses Fermilab's NuMI
$\nu_\mu/\bar{\nu}_\mu$ beamline.  Its detector is a 14 kton totally active scintillator detector (TASD), placed at a distance of 810 km
from Fermilab,  near Ash River, which is $0.8^\circ$ off-axis from the NuMI beam. It also has a 0.3 kton near detector  located at the Fermilab site 
to monitor the unoscillated neutrino or anti-neutrino flux.
It has already started data taking  from late 2014. The experiment is scheduled to have three years run in neutrino mode followed by three
years run in anti-neutrino mode with a NuMI beam power of 0.7 MW and 120 GeV proton energy, corresponding to $6\times 10^{20}$
p.o.t per year.  Apart from the precise measurement of $\theta_{13}$ and the atmospheric parameters, 
it aims to determine the unknowns such as neutrino mass ordering,
leptonic CP-violation,  and the octant of $ \theta_{23}$ by the measurement of
$\nu_\mu/\bar{\nu}_\mu \to \nu_e/ \bar{\nu}_e$ oscillations.

For the simulation of NO$\nu$A experiment, the detector properties and other necessary details are taken from  \cite{3,sdr}. 
We have used the following input true values of neutrino oscillation parameters in our simulations: $|\Delta m_{eff}^2| = 2.4 \times 10^{-3}~ {\rm eV}^2$,
$\Delta m_{21}^2 = 7.6 \times 10^{-5}~ {\rm eV}^2 $, $\delta_{CP}=0$,  $\sin^2\theta_{12} = 0.32$, $\sin^2 2\theta_{13} = 0.1$ and $\sin^2\theta_{23} =$ 0.5.
The relation between the atmospheric parameter $\Delta m_{eff}^2$ measured in MINOS and  the standard  oscillation parameter $\Delta m_{31}^2$ in nature
is given as \cite{4}
\begin{equation}
\Delta m_{31}^2 = \Delta m_{eff}^2 +\Delta m_{21}^2 (\cos^2 \theta_{12}- \cos \delta_{CP} \sin \theta_{13} \sin 2\theta_{12} \tan \theta_{23})\;,
\end{equation}
where $\Delta m_{eff}^2$ is taken to be positive for Normal Ordering (NO) and negative for Inverted Ordering (IO).

\begin{figure}[!htb]
\includegraphics[width=7.5cm,height=6cm]{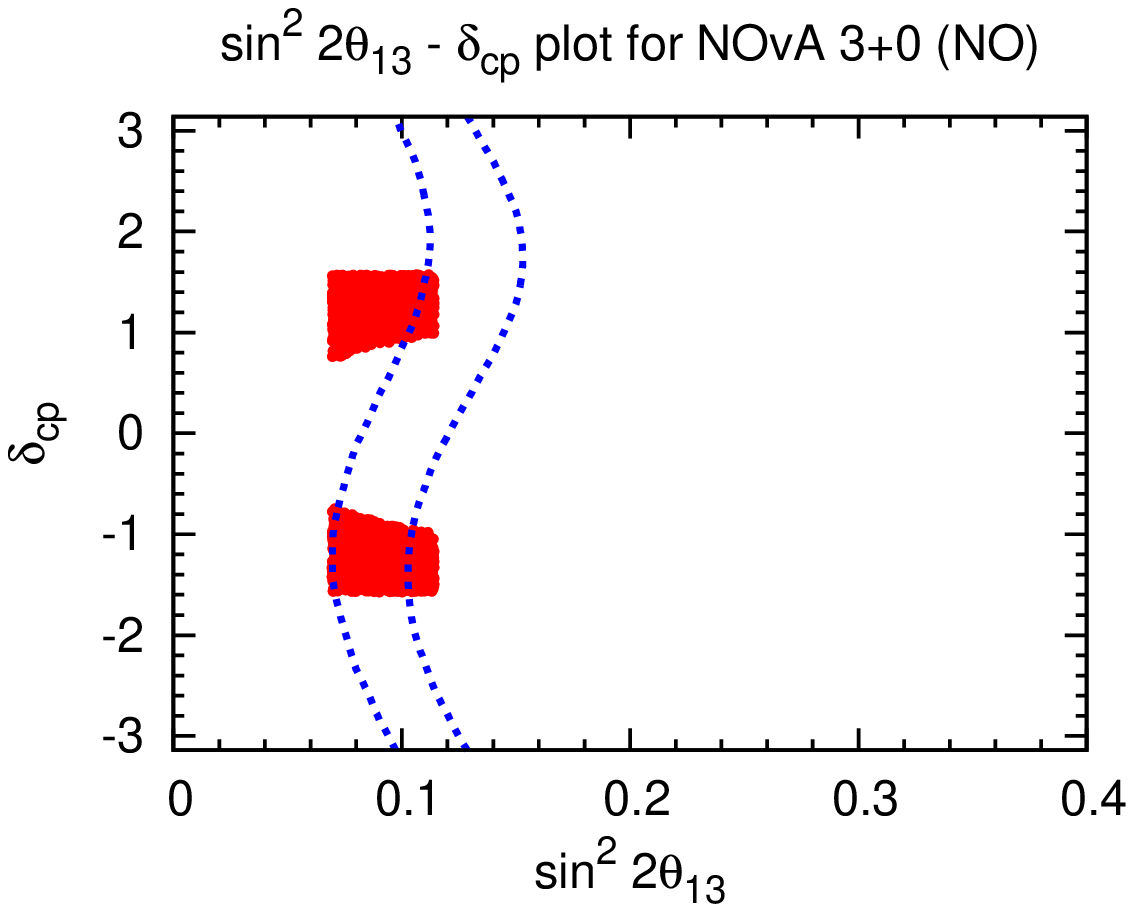}\\
\includegraphics[width=7cm,height=6cm]{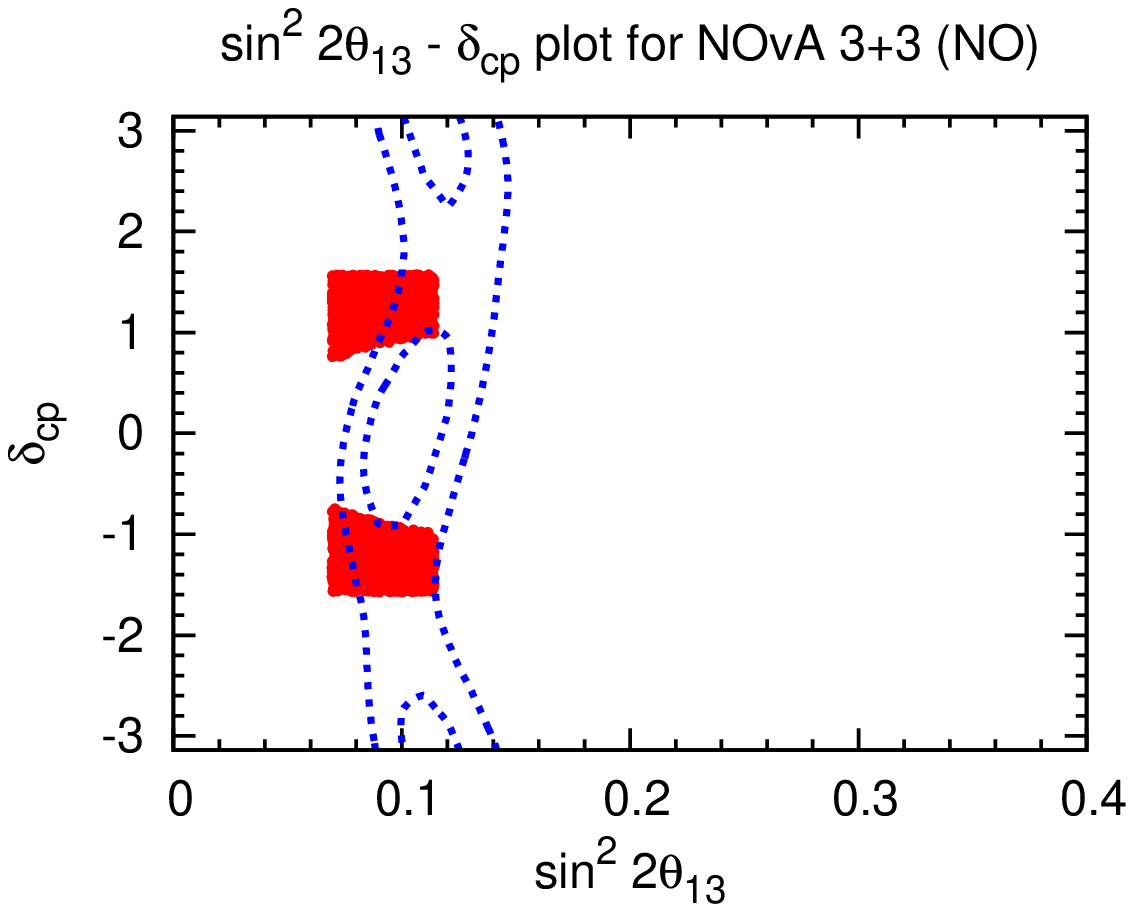}
\includegraphics[width=7cm,height=6cm]{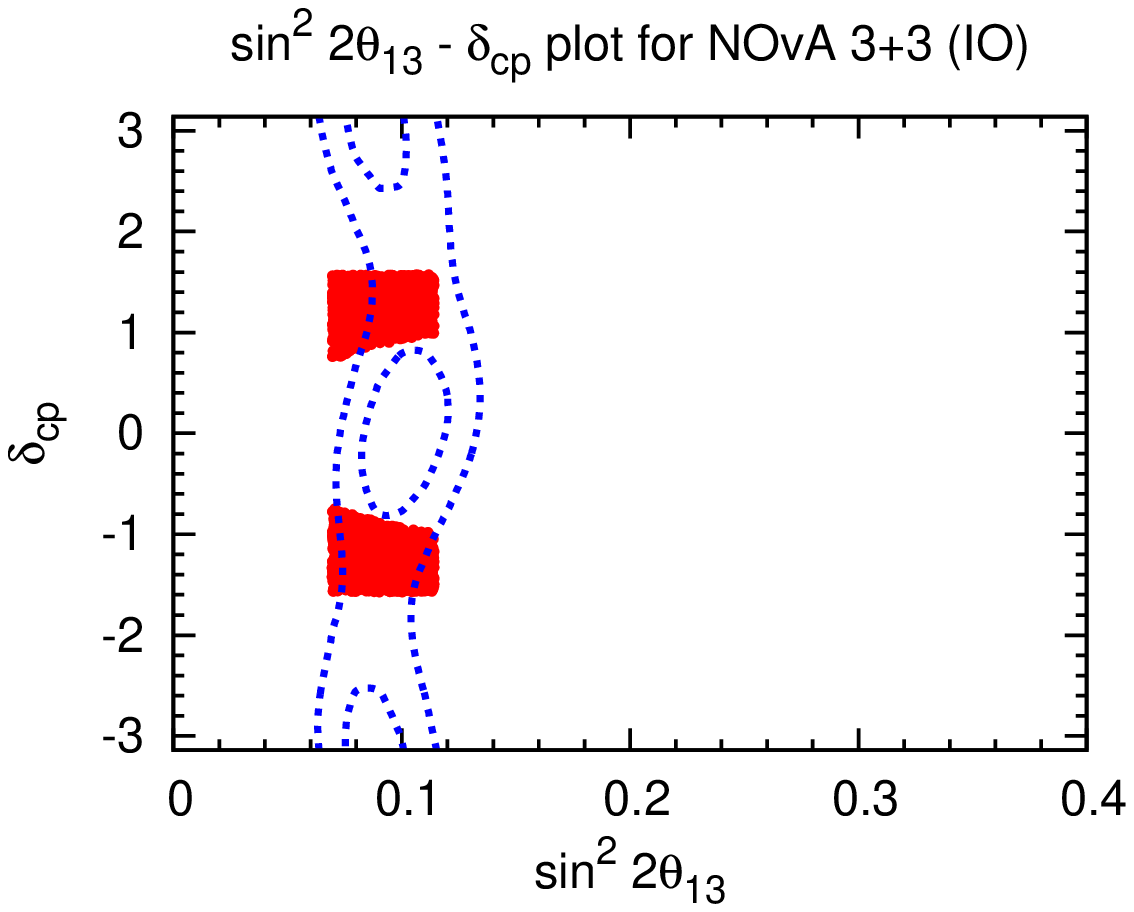}
\caption{The correlation between $\delta_{CP}$ and $\sin^2 2\theta_{13}$ for TBM mixing pattern (red regions) superimposed on
expected NO$\nu$A data where the blue dashed lines (top panel) represent the 1$\sigma$ contours for 3 years of neutrino data taking
with NO as test ordering, the blue lines in the  bottom-left (NO as test ordering) and bottom-right (for IO as test ordering) panels represent the
1$\sigma$ and $3\sigma$ contours for (3$\nu$+3$\bar \nu$)
years of running.}
\end{figure}

In order to obtain the allowed region for $\sin^22\theta_{13}$ and  $\delta_{CP}$, we  generate the  true event spectrum by keeping the 
above mentioned neutrino oscillation parameters as true values and generate the  test event spectrum by varying the  test values of  $\sin^22\theta_{13}$ in the
range [0.02:0.25] and that of $\delta_{CP}$ in its full range [$-\pi:\pi$]. Finally, we calculate $\Delta\chi^2$ by comparing the true  and
test event spectra. The obtained results in the $\sin^2 2 \theta_{13}-\delta_{CP}$ plane are shown in Figure-2, which are overlaid by our
predicted value of $\delta_{CP}$. The top panel shows the $1\sigma$ contours for the running of ($3\nu + 0 \bar{\nu}$) years, with NO as the true
hierarchy. The bottom left (right) panel represents ($3\nu + 3 \bar{\nu}$) years of data taking with NO (IO) as the true hierarchy. In these plots,
the inner regions (bubbles) correspond to $1\sigma$ contours whereas the outer  curves represent $3\sigma$ contours. From these plots,
one can see that our results are  supported by NO$\nu$A data within $3\sigma$ C.L., however, with (3$\nu$+3$\bar \nu$) years of data taking, NO$\nu$A could
marginally exclude these results at $1\sigma$ C.L.

Next we would like to briefly mention about the implications of future generation long baseline experiments such as Hyder-Kamiokande (T2HK) and Deep Underground Neutrino 
Experiment (DUNE) experiments in our predicted results. All the details for simulation of T2HK experiment are taken from
\cite{sdr} for (3$\nu$+7$\bar \nu$) years of running. The DUNE experiment which is basically  slightly upgraded version of LBNE experiment, plans to use a 40 kton Liquid Argon detector.
Except the detector volume other characteristics  are taken  from \cite{sdr1} for the simulation
for  (5$\nu$+5$\bar \nu$) years of  data taking. We use the same true values of other input parameters 
as done for NO$\nu$A experiment. The  correlation plots between $\delta_{CP}$ and $\sin^2 2 \theta_{13}$ are shown in Figure-3,  overlaid by  our
predicted values for TBM. The plots on the top (bottom)  panel are for DUNE (T2HK) experiment with NO/IO as the true
ordering as labeled in the plots. It can be seen from these figures that as
 the $\delta_{CP}-\sin^2 2 \theta_{13}$ parameter space is severely constrained, our predicted results are expected to
be precisely verified by these experiments.  

\begin{figure}[!htb]
\includegraphics[width=7cm,height=6cm]{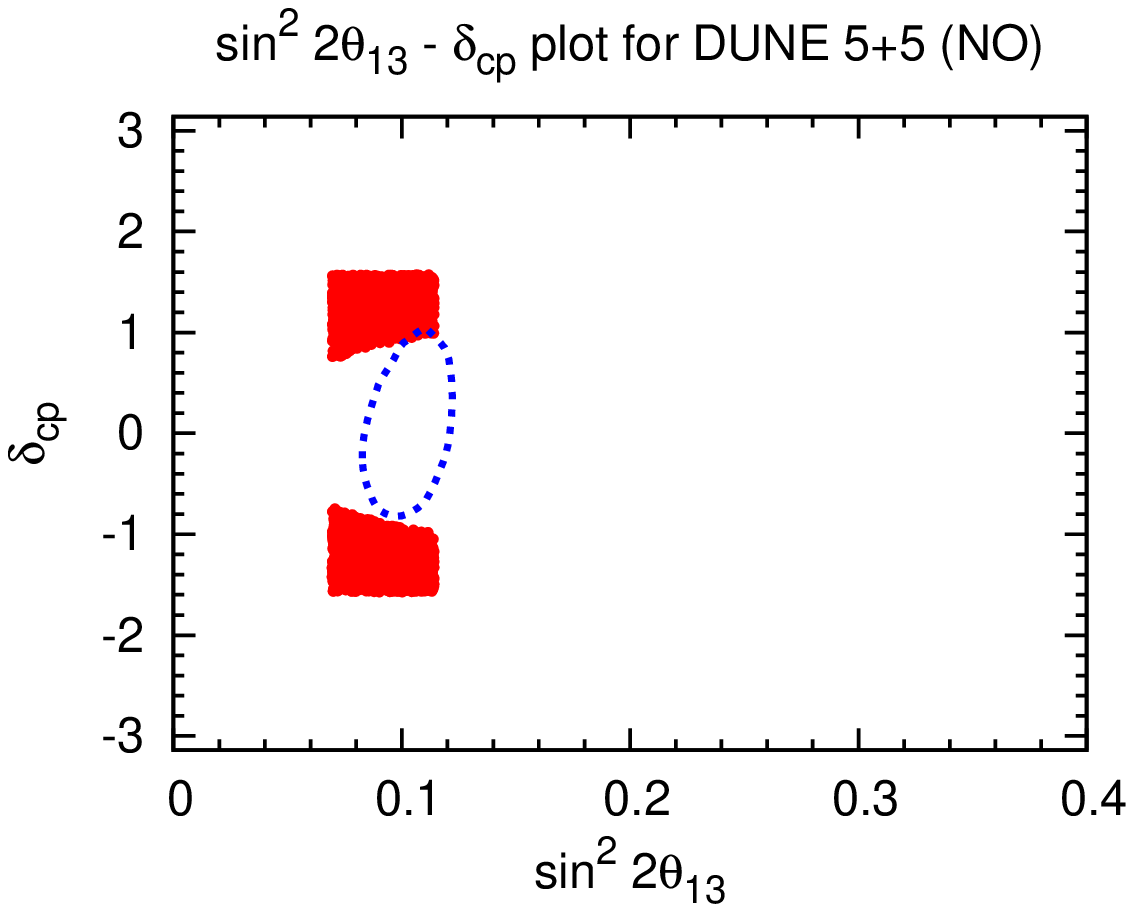}
\includegraphics[width=7cm,height=6cm]{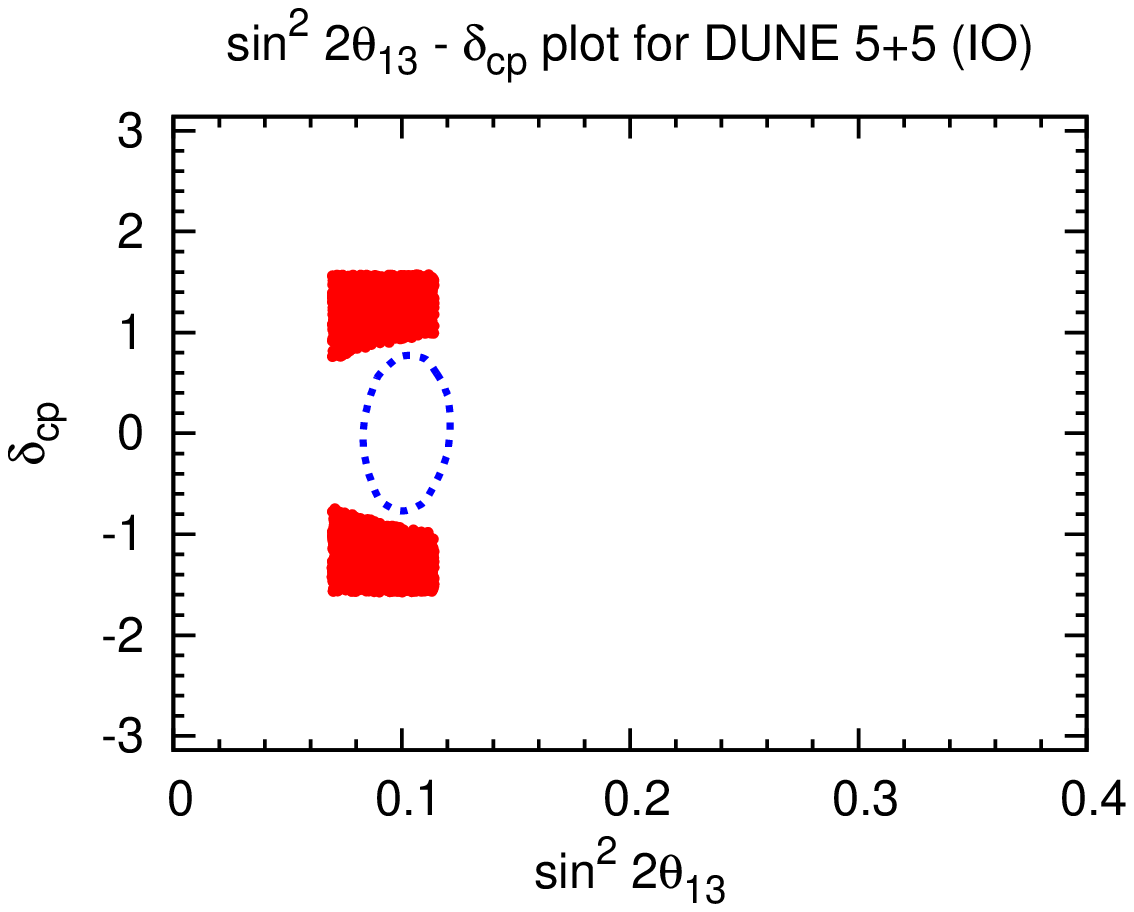}
\includegraphics[width=7cm,height=6cm]{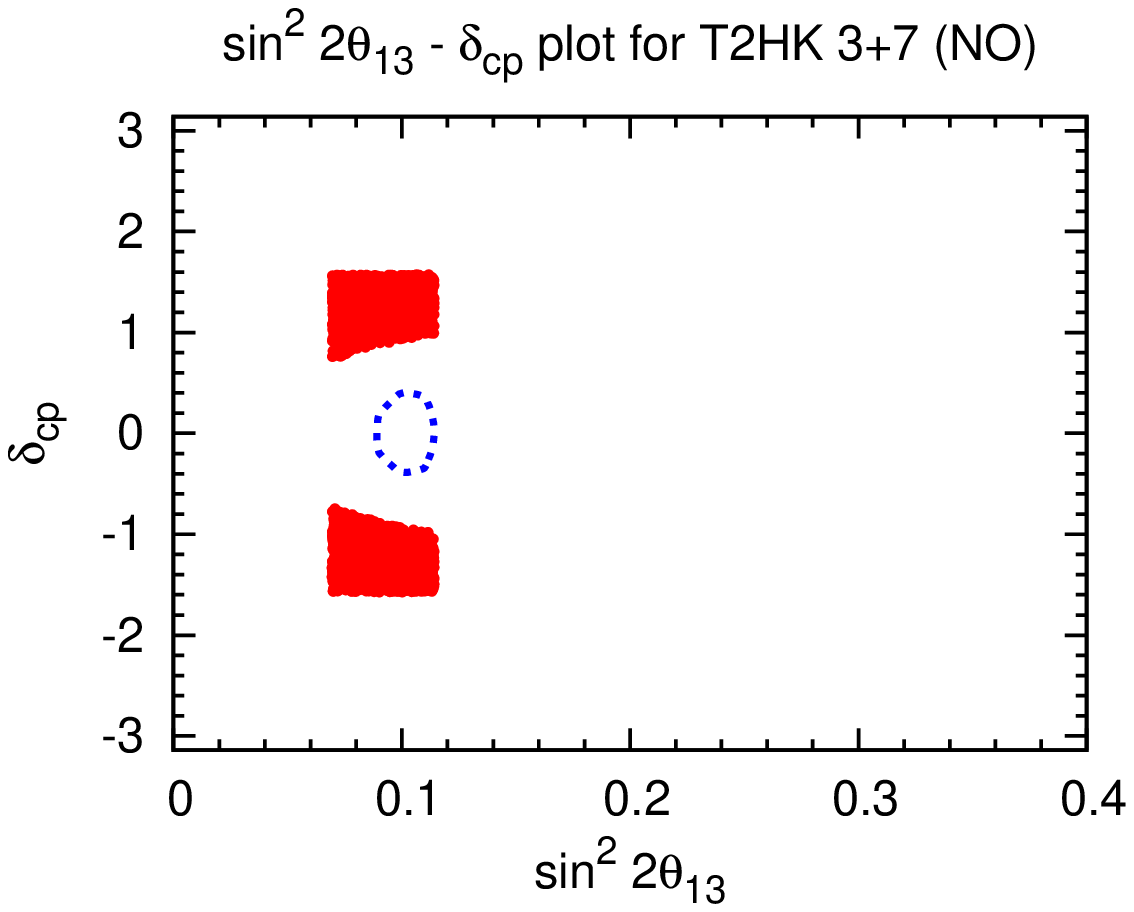}
\includegraphics[width=7cm,height=6cm]{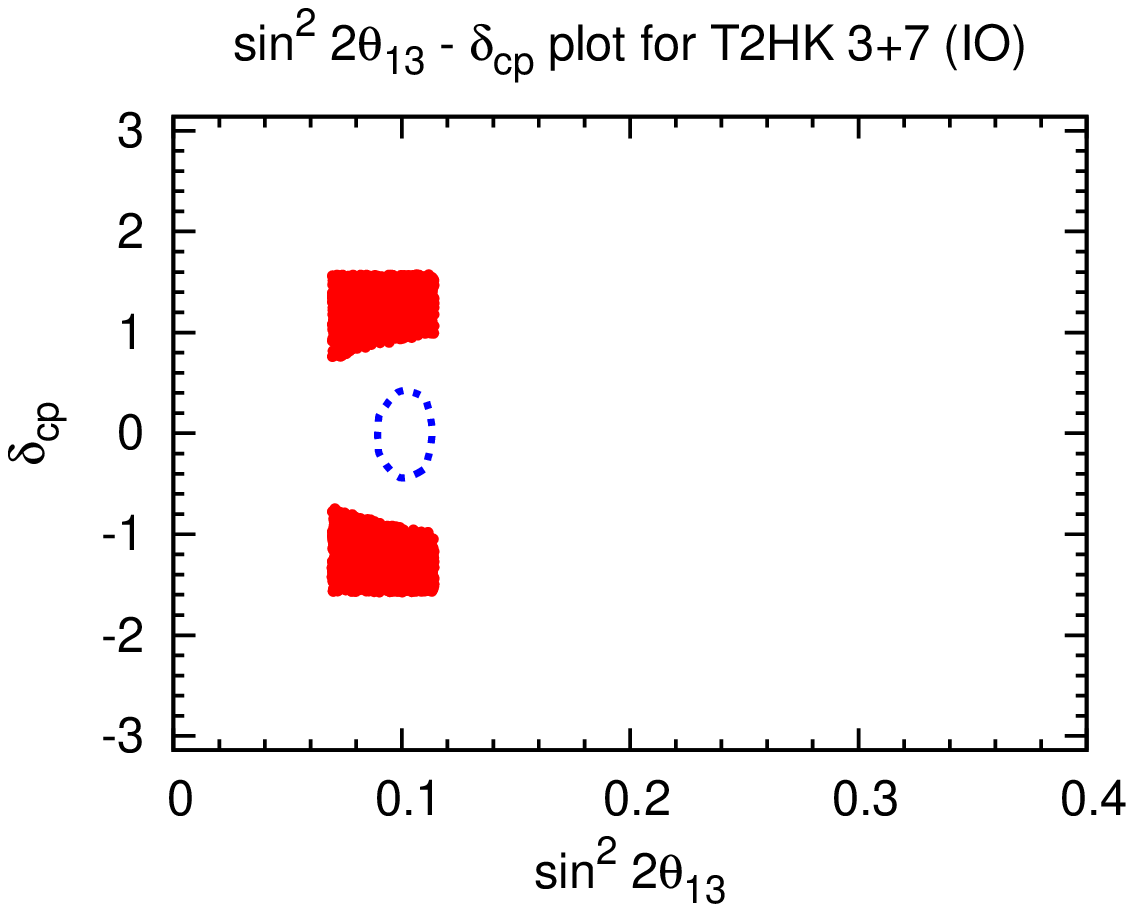}
\caption{The correlation between $\delta_{CP}$ and $\sin^2 \theta_{13}$ for TBM mixing pattern (red regions) superimposed on
expected DUNE data (top panels) where the blue dashed lines represent the 3$\sigma$ contours for  (5$\nu$+5$\bar \nu$) years of  data taking,
while the bottom panels represent the T2HK results for  (3$\nu$+7$\bar \nu$)
years of running.}
\end{figure}

\begin{table}[htb]
\begin{center}
\vspace*{0.1 true in}
\begin{tabular}{|c|c|c|c|}
\hline
 Deviation type & Neutrino mixing  & $\delta_{CP}$ Range & $|J_{CP}|$ Range  \\
& matrix pattern & (in radian)&\\

\hline
23 rotation to $U_\nu^0$ &  TBM and GRB & $  \pm(0.7 - 1.5) $ & $(0.02 - 0.04)$ \\

\hline

13 rotation to $U_\nu^0$ &  TBM,  GRA and  GRB  & $ \pm(0 - 1.5) $ & $(0 - 0.04)$ \\
\hline
12 and 13 rotation to $U_l$ &  TBM and GRB & $\pm(1.2 - 1.55) $ & $(0.03 - 0.04)$ \\
& GRA  & $\pm(0.6 - 1.5)  $ & $ (0.02 -0.04)$ \\
& HG  & $\pm(0 - 1.3)  $ & $ (0 -0.035)$ \\
& BM & $\pm(0 - 0.8)  $ & $(0 -0.03)$ \\

\hline
\end{tabular}
\end{center}
\caption{Predicted range of the CP phase $\delta_{CP}$ and the Jarlskog invariant $|J_{CP}|$ due to possible deviations
for various neutrino mixing patterns.}
\end{table}
\subsection{Deviation due to 13 rotation}
Next we consider the corrections arising from an additional (13) rotation in the neutrino sector for which the rotation matrix can be given as
\be
U_{PMNS} = U_\nu^0   \left ( \begin{array}{ccc}
 \cos \phi    & 0 & e^{-i  \alpha} \sin \phi \\
0    & 1    & 0 \\
 -e^{i \alpha} \sin \phi  & 0   & \cos \phi\\
\end{array}
\right )
\ee
Proceeding in the similar way as done in the previous case, we obtain the mixing angles using Eq. (\ref{mixing1}) as
\bea
\sin \theta_{13} &=& \cos \theta_{12}^\nu \sin \phi\;, \label{res-c}\\
\sin^2 \theta_{12} &=& \frac{\sin^2 \theta_{12}^\nu}{1- \cos^2 \theta_{12}^\nu \sin^2 \phi}\;, \label{res-cd}\\
\sin^2 \theta _{23} &=&  \frac{1}{2}\left [ \frac{\cos^2 \phi + \sin  \theta_{12}^\nu \sin 2 \phi \cos \alpha + \sin^2 \theta_{12}^\nu   \sin^2 \phi}
{1- \cos^2 \theta_{12}^\nu  \sin^2 \phi}\right ]\;.
\eea
Analogously, the Jarlskog invariant and the CP violating phase $\delta_{CP}$ are  given as
\bea
J_{CP} = - \frac{1}{8} \cos \theta_{12}^\nu \sin 2 \theta_{12}^\nu \sin 2 \phi \sin \alpha \;,
\eea
and
\be
\sin \delta_{CP}= -\frac{(1- \cos^2 \theta_{12}^\nu \sin^2 \phi)\sin \alpha}{\Big[(1-\cos^2 \theta_{12}^\nu \sin^2 \phi)^2 - \sin^2 \theta_{12}^\nu
\sin^2 2 \phi \cos^2 \alpha \Big]^{1/2}}\;.
\ee
In this case one obtains from Eqs. (\ref{res-c}) and  (\ref{res-cd})
\be
\sin^2 \theta_{12}= \frac{\sin^2\theta_{12}^\nu}{1-\sin^2 \theta_{13}},
\ee
which implies that $\sin^2 \theta_{12} > \sin^2 \theta_{12}^\nu$. This in turn implies that BM and HG mixing patterns cannot accommodate
the observed value of $\theta_{12}$ within its $3\sigma$ range.

From Eq. (\ref{res-c}) and using the $3 \sigma$ allowed range of $\sin^2 \theta_{13}$ the allowed range of $\phi$ is found to be in
the range $(9-15)^\circ$ for various mixing patterns.
Now using this value of $\phi$ and  varying the free phase parameter $ \alpha $  in the range $-\pi \leq \alpha \leq   \pi$, 
we obtain the correlation plots between
$\sin^2 \theta_{12}$ and $\sin^2\theta_{13}$ as shown in the top left panel of Figure-3, where red, blue and green  plots are for
TBM, GRB and GRA mixing patterns. The correlation plot for HG and BM  forms are not shown in the figure as they lie outside the allowed $3 \sigma$
region of $\sin^2 \theta_{12}$. The  $\delta_{CP}$ phase is very loosely constrained  in this case as presented in Figure-4. We also overlaid the
predicted value of $\delta_{CP}$ for TBM over the NO$\nu$A simulated data. In this case also the predicted result is
consistent with expected NO$\nu$A data. The correlation plots between $\delta_{CP}$ and $\sin^2  \theta_{23}$,  $J_{CP}$ and 
$\sin^2  \theta_{13}~ (\sin^2  \theta_{23})$, 
as well as between  $J_{CP}$ and $\delta_{CP}$ 
are also shown in the figure.
From the plots it can be seen that
it could be possible to have large CP violation ${\cal O}(10^{-2})$ in the lepton sector.

It should be noted that for deviation due to 12 rotation matrix does not accommodate the observed value of $\theta_{13}$ as $U_{e3}=0$
for such case. 
\begin{figure}[!htb]
\includegraphics[width=6.0cm,height=4.7cm]{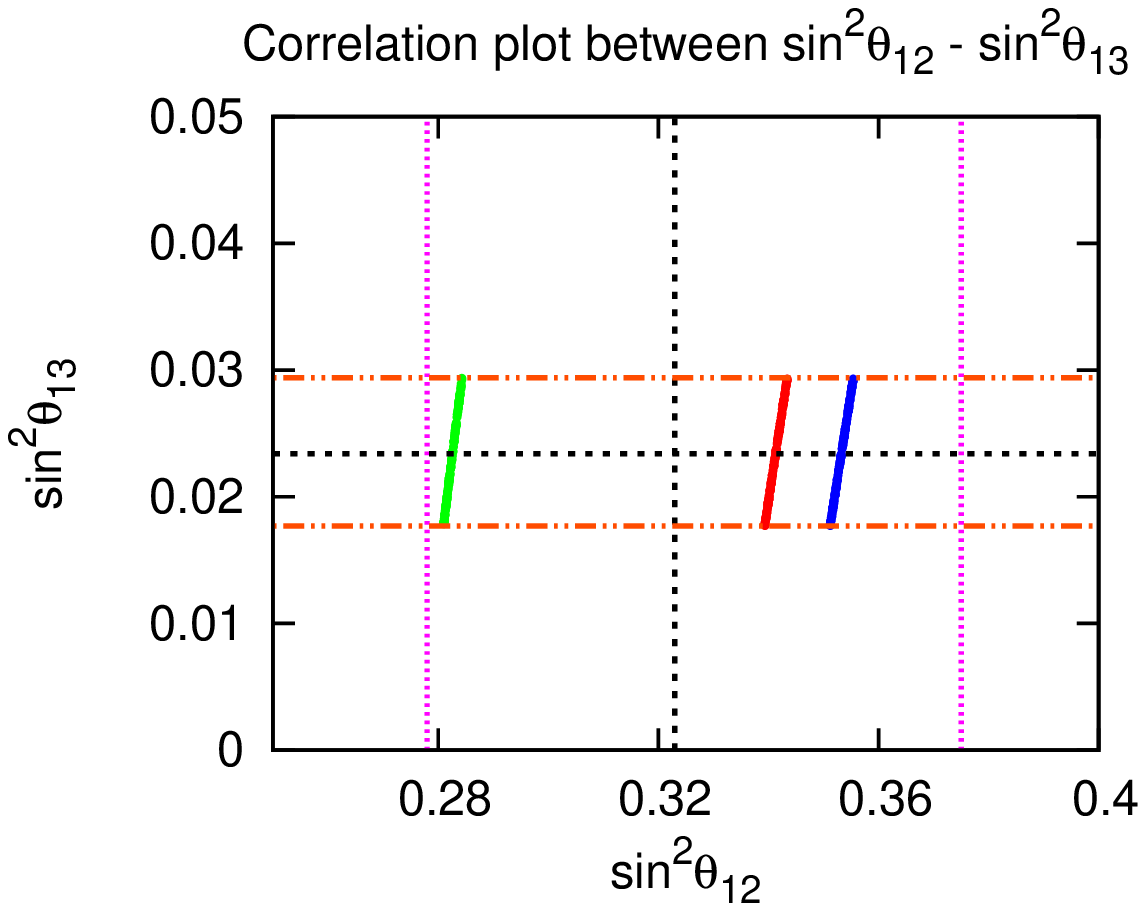}
\includegraphics[width=6.0cm,height=4.7cm]{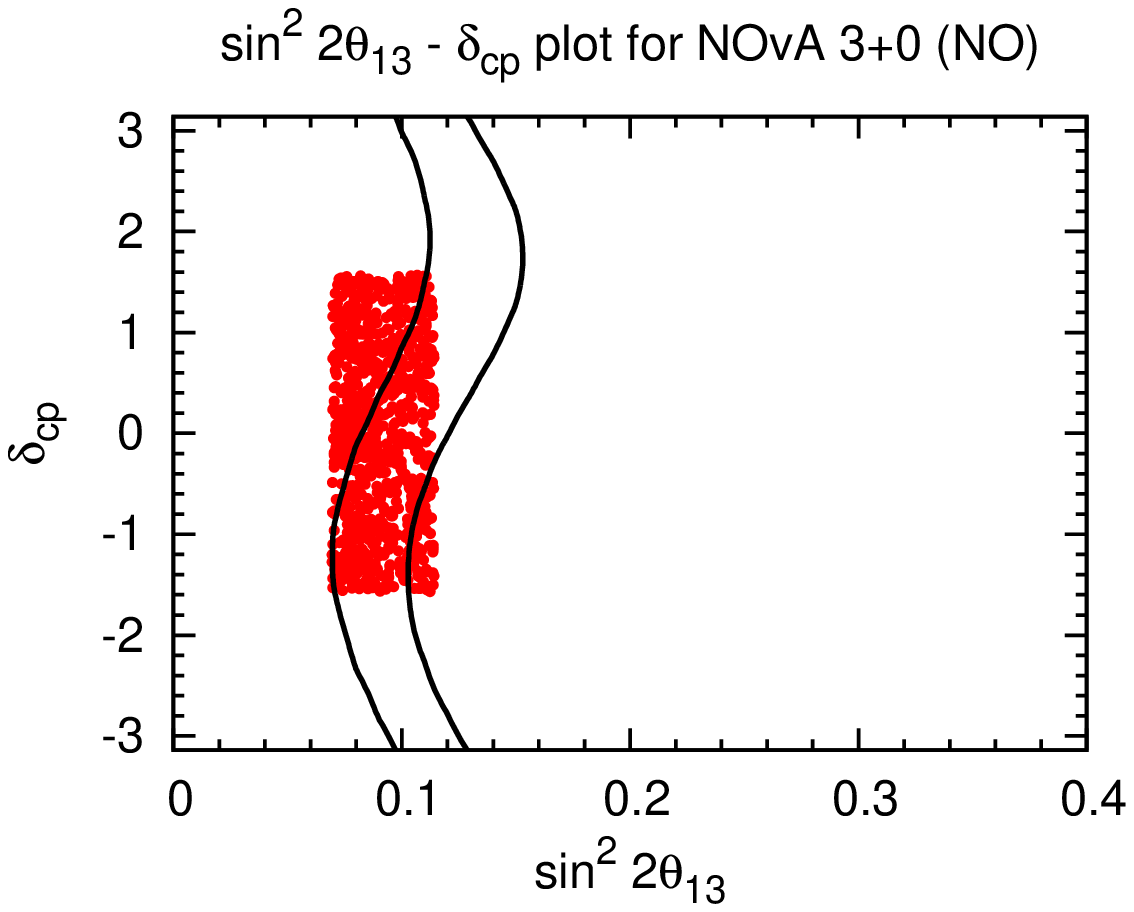}
\includegraphics[width=6.0cm,height=4.7cm]{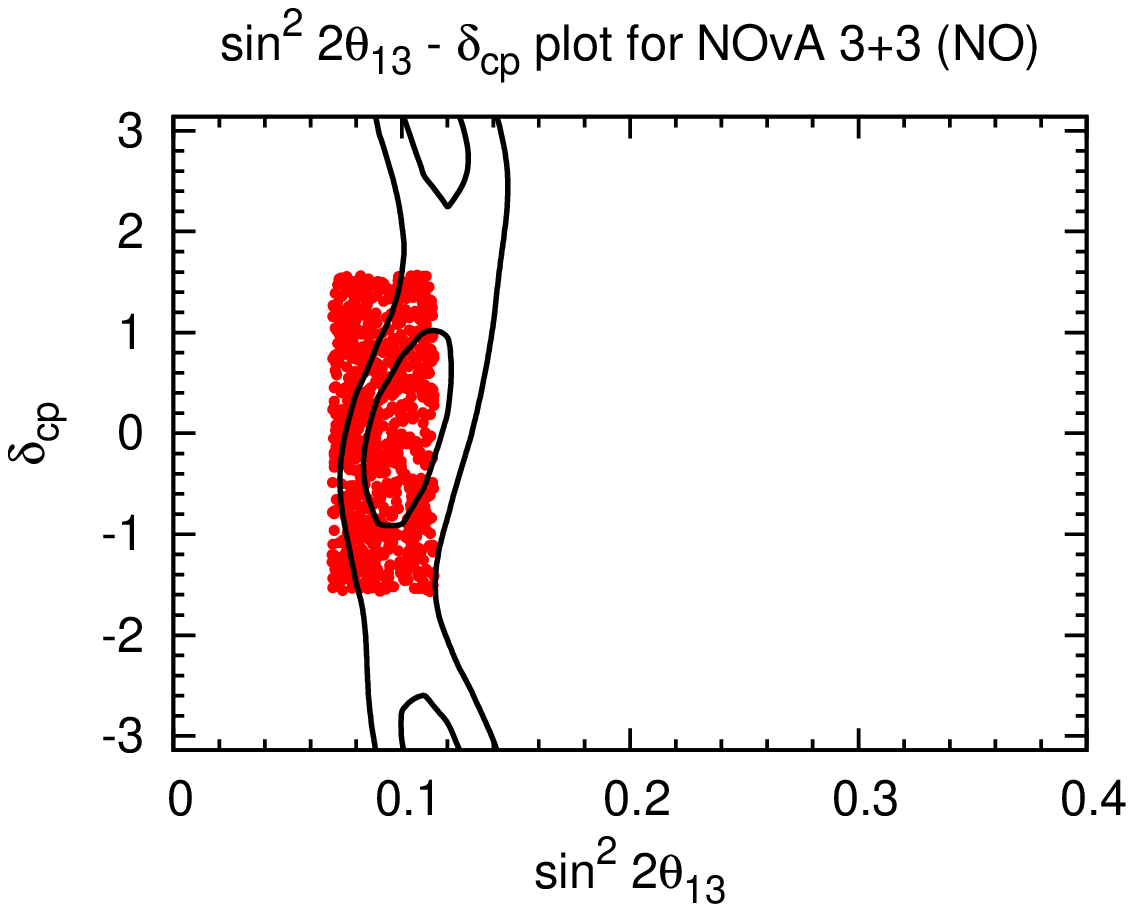}
\includegraphics[width=6.0cm,height=4.7cm]{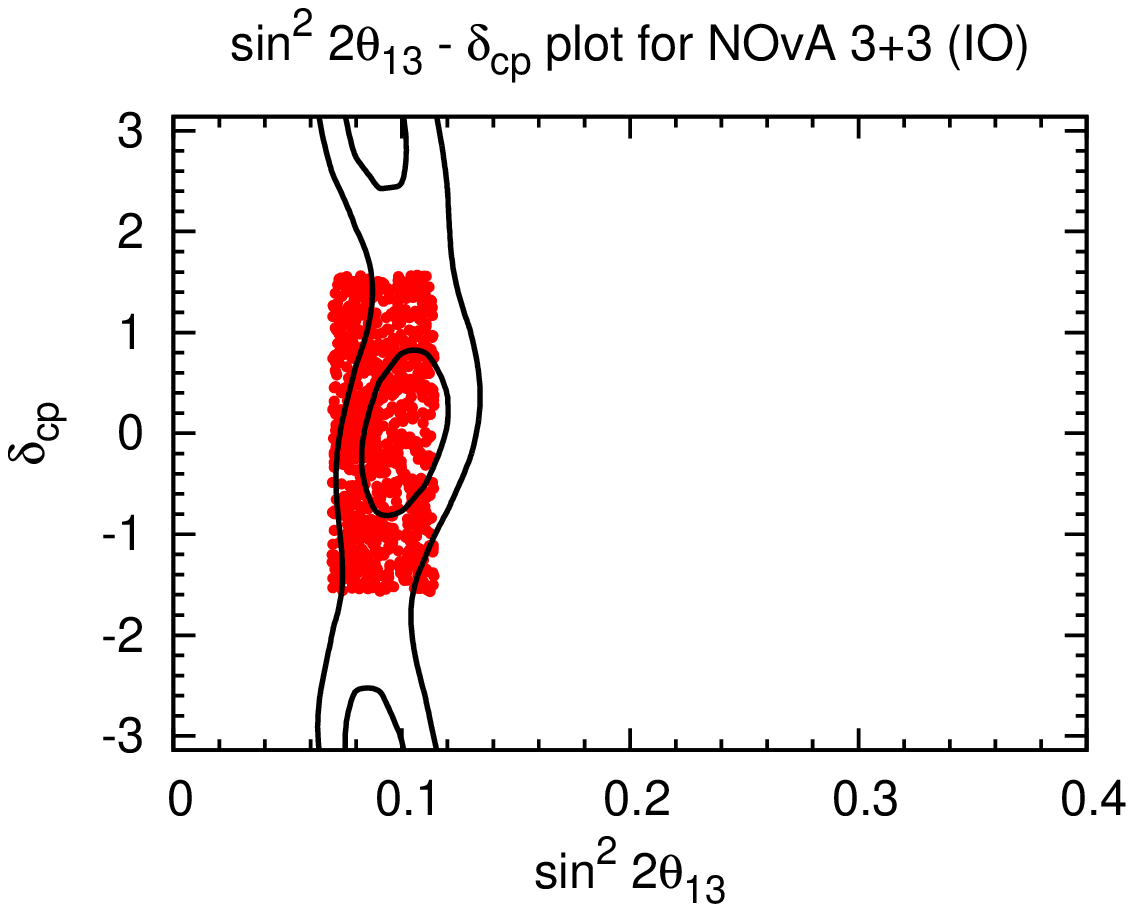}
\includegraphics[width=6.0cm,height=4.7cm]{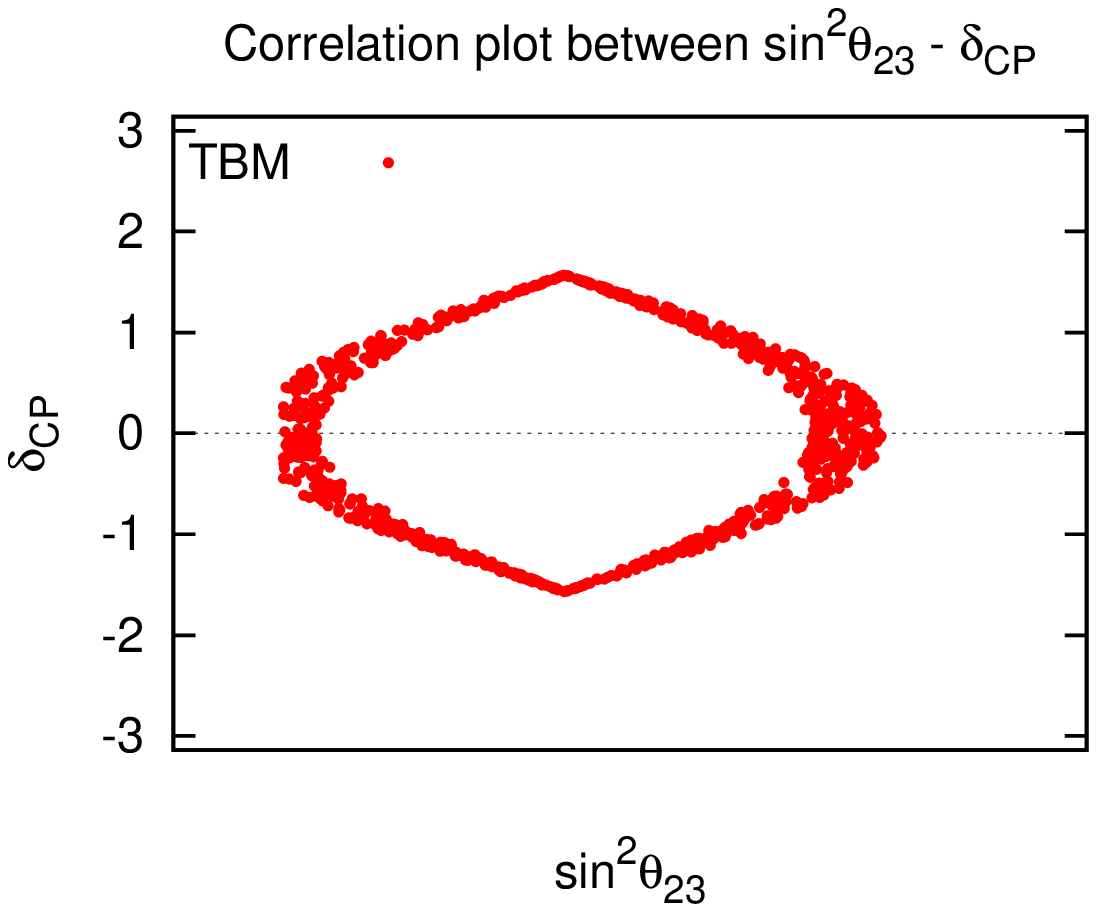}
\includegraphics[width=6.0cm,height=4.7cm]{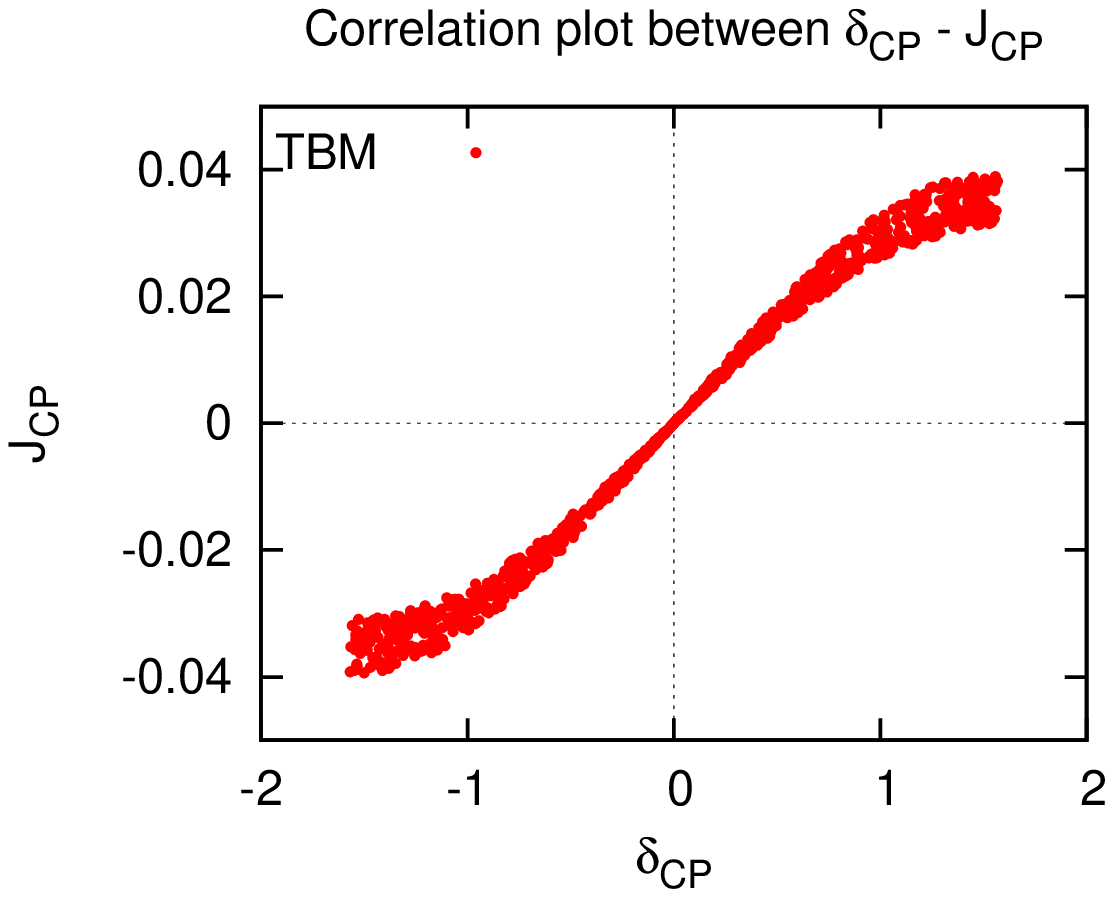}
\includegraphics[width=6.0cm,height=4.7cm]{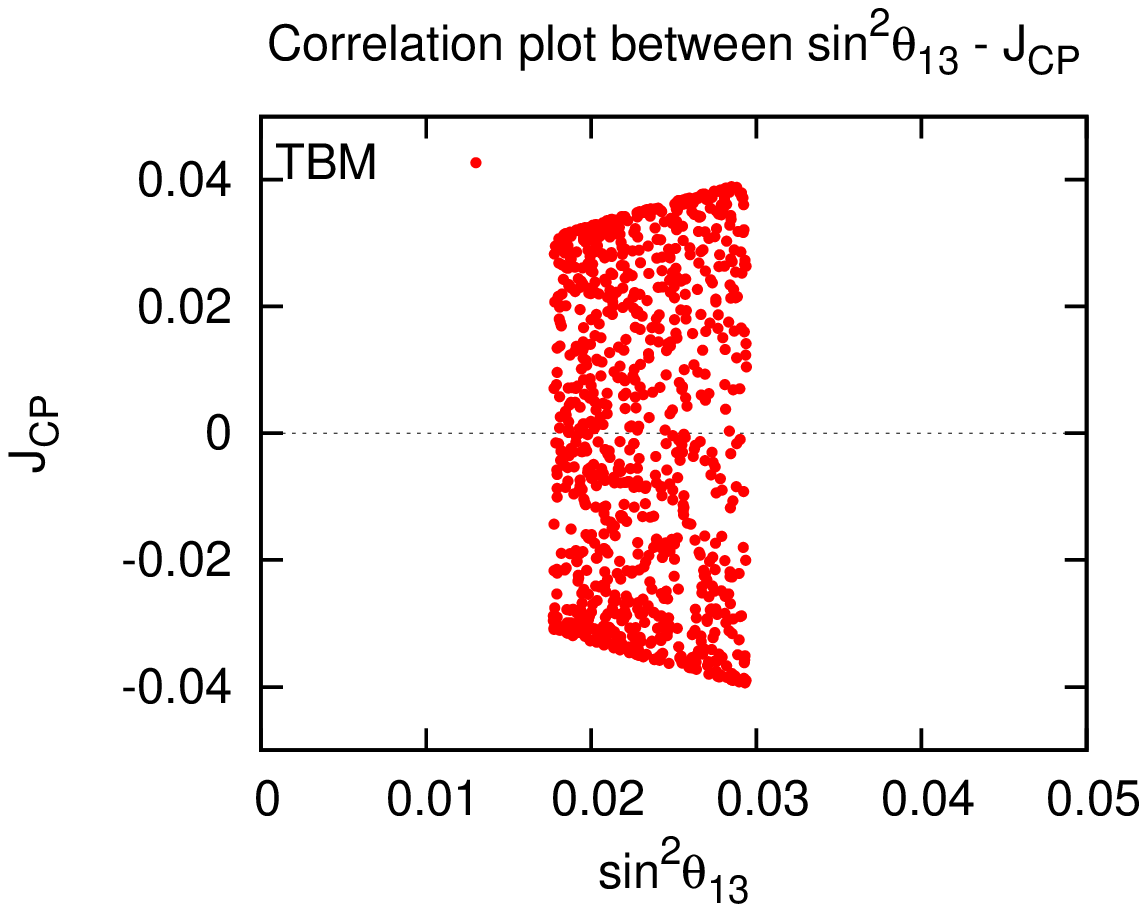}
\includegraphics[width=6.0cm,height=4.7cm]{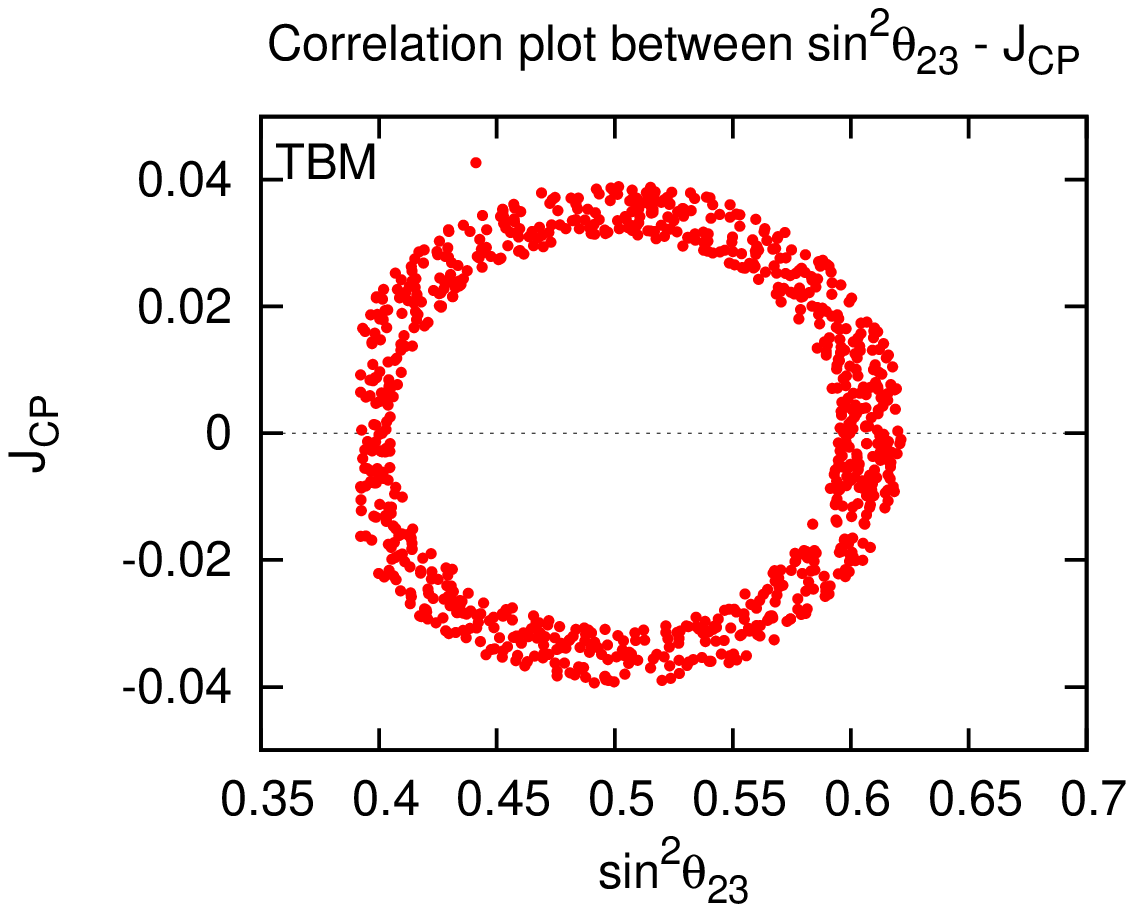}
\caption{Correlation plots between different oscillation parameters  due to 13 deviation in the neutrino sector. In the top left panel
the red, blue and  green  plots correspond to TBM, GRB and GRA  mixing patterns. Other plots represent the
correlation between different mixing parameters as indicated in the plot labels for TBM mixing pattern.
The black solid lines (in the top right and second  panels) represent the expected experimentally allowed parameter space (same as the dotted
blue lines in Fig. 2).}
\end{figure}

\section{Deviation in the charged lepton sector}

In this section we will consider the deviation arising in the charged lepton sector.
For the study of lepton mixing  it is generally assumed that the charged lepton mass matrix is diagonal
and hence, the corresponding  mixing matrix as an identity
matrix. The deviation  in the charged lepton sector and its possible
consequences  have been studied by various authors \cite{charged-lepton, charged-lepton1, petcov, petcov1}. 
In Refs. \cite{petcov,petcov1}, the form for $U_l$ is considered to be product of two
orthogonal matrices describing rotations in 23 and 12 planes, which corresponds to two possible orderings,
`standard' with $U_l \propto R_{23}(\theta_{23}^l)R_{12}(\theta_{12}^l)$ and `inverse'
with $U_l \propto R_{12}(\theta_{12}^l)R_{23}(\theta_{23}^l)$. 
Using these forms for the lepton mixing matrix 
the values of $\delta_{CP}$ and the rephasing invariant
$J_{CP}$ have been predicted for the cases TBM, BM, LC, GRA, GRB and HG forms of 
neutrino mixing matrix $U_\nu$.  
They have obtained the  predictions  for $\delta_{CP}$  as $\delta_{CP} \simeq \pi $ for BM (LC) and $\delta_{CP} \simeq
3 \pi/2$ or $\pi/2$ for TBM, GRA, GRB and HG. 
Here, we  consider the simplest case where 
 the deviation matrix can be represented  as a single rotation matrix in the ($ij$) plane, as done in the previous section
for the  neutrino sector.

Now  considering the   deviation to the charged lepton mixing
matrix as a  unitary rotation matrix either in (12), (23) or (13) plane,
one can write the PMNS matrix as
\be
U_{PMNS}=U_{ij}^\dagger U_\nu^0\;,
\ee
where $U_{ij}$ is the rotation matrix in $(ij)$ plane and $U_\nu^0$ is any one of  the standard neutrino mixing matrix  form TBM/BM/GRA/GRB/HG.
However, corrections arising due to $U_{23}$ rotation matrix is ruled out as it gives vanishing $U_{e3}$.

\subsection{Deviation due to rotation in 12 and 13 sector}

Including the additional correction matrix $U_{12}$ to the charged lepton sector, one can write the PMNS matrix as
\be
U_{PMNS} =   \left ( \begin{array}{ccc}
 \cos \phi    &  -e^{-i  \alpha} \sin \phi & 0 \\
 e^{i \alpha} \sin \phi  &  \cos \phi &0\\
0    & 0    & 1 \\
\end{array}
\right ) U_\nu^0\;.
\ee
In this case we get the mixing angles as
\bea
\sin \theta_{13} &=&  \frac{\sin \phi}{\sqrt 2}\;,\label{res-d}\\
\sin^2 \theta_{12} &=& \frac{2 \sin^2 \theta_{12}^\nu \cos^2 \phi + \cos^2 \theta_{12}^\nu \sin^2 \phi - \frac{1}{\sqrt 2} \sin  2
\theta_{12}^\nu \sin 2 \phi  \cos \alpha}{1+ \cos^2 \phi}\;,\\
\sin^2 \theta_{23} &=& \frac{\cos^2 \phi}{1+\cos^2 \phi}\;.\label{res-da}
\eea
With Eqs. (\ref{res-d}) and (\ref{res-da}), we obtain the relation
\be
\sin^2 \theta_{23}= 1- \frac{1}{2 \cos^2 \theta_{13}}\;,
\ee
which implies that $\sin^2 \theta_{23}<1/2$.
The Jarlskog invariant in this case  is found to be
\be
J_{CP}= - \frac{1}{8 \sqrt 2} \sin 2 \theta_{12}^\nu \sin 2 \phi \sin \alpha\;,
\ee
and the CP violating phase as
\be
\sin \delta_{CP} = - \frac{(1+\cos^2 \phi) \sin 2 \theta_{12}^\nu \sin \alpha }{2 \sqrt Y}\;,
\ee
where
\bea
Y&=& \Big(2 \sin^2 \theta_{12}^\nu  \cos ^2 \phi + \cos^2 \theta_{12}^\nu \sin^2 \phi - \frac{1}{\sqrt 2} \sin 2 \phi \sin 2
\theta_{12}^\nu \cos \alpha \Big)\nn\\
&\times & \Big(1+ \cos 2 \theta_{12}^\nu \cos^2 \phi - \cos^2 \theta_{12}^\nu \sin^2 \phi + \frac{1}{\sqrt 2} \sin 2
\theta_{12}^\nu \sin 2 \phi \cos \alpha \Big )\;.
\eea
Proceeding in a similar fashion as in the previous cases and  considering the $3 \sigma$ allowed range of
$\theta_{13}$, one can obtain the allowed range of $\phi$ with Eq. (\ref{res-d})  
as  $(10-15)^\circ $. 
Now varying the free parameters $\phi$ and $\alpha$ in their allowed ranges, we obtain the correlation plots between various mixing parameters as
depicted in Figure-5. It should be noted that the correlation plots between $\sin^2 \theta_{13}$ and $\sin^2 \theta_{23}$ remain same
for all the forms of neutrino mixing matrix $U_\nu^0$  as these mixing angles depend only on the free parameter $\phi$ and
are independent of $\theta_{12}^\nu$ (which takes different values for different mixing patterns).
For the correlation plots between $\delta_{CP}-\sin^2 2 \theta_{13}~(\sin^2 \theta_{12})$ and $J_{CP}-\sin^2 \theta_{13}$,
the red, green, blue and magenta regions correspond to TBM, GRA, HG and BM mixing patterns. The GRB mixing pattern
predicts the same constraints as TBM pattern and hence, the corresponding results are not shown in the plots. 
 Furthermore, the CP violating phase is severely constrained in this scenario and the
Jarlskog invariant is found to be significantly large as seen from the figure.

Next we consider deviation due to additional rotation in 13 sector. In this case the PMNS matrix is given as
\be
U_{PMNS} =   \left ( \begin{array}{ccc}
 \cos \phi    & 0 & -e^{-i  \alpha} \sin \phi \\
0    & 1    & 0 \\
 e^{i \alpha} \sin \phi  & 0   & \cos \phi\\
\end{array}
\right ) U_{\nu}^0
\ee
The mixing angles obtained are
\bea
\sin \theta_{13} &=& \frac{\sin \phi}{\sqrt 2}\;, \nn\\
\sin^2 \theta_{23} &=& \frac{1}{1+ \cos^2 \phi}\;,\nn\\
\sin^2 \theta_{12} &=& \frac{2 \sin^2 \theta_{12}^\nu \cos^2 \phi + \cos^2 \theta_{12}^\nu \sin^2 \phi
- \frac{1}{\sqrt 2} \sin 2 \theta_{12}^\nu \sin 2 \phi \cos \alpha}{1+\cos^2 \phi}\;.
\eea
In this case we obtain 
\be
\sin^2 \theta_{23}=\frac{1}{2 \cos^2 \theta_{13}}\;,
\ee
which implies $\sin^2 \theta_{23} > 1/2$.
\begin{figure}[!htb]
\includegraphics[width=7cm,height=6cm]{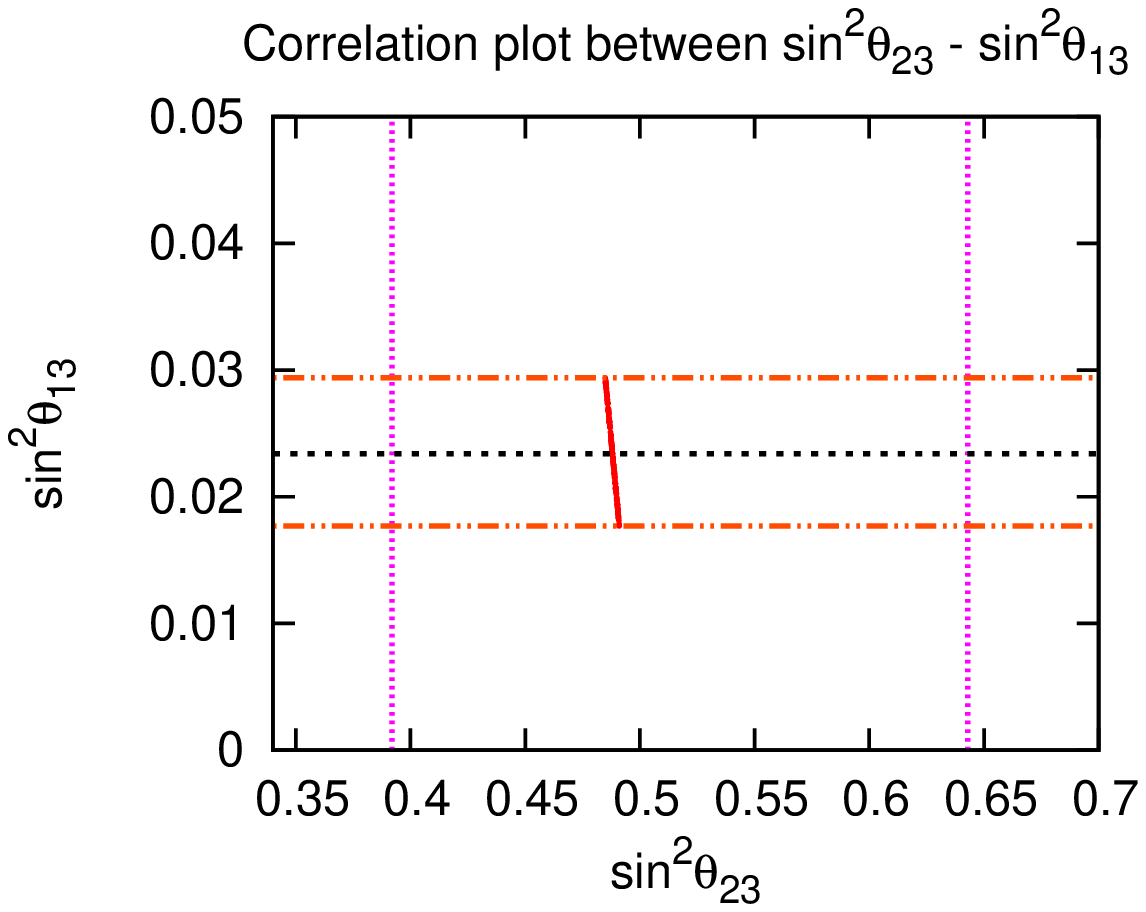}
\includegraphics[width=7cm,height=6cm]{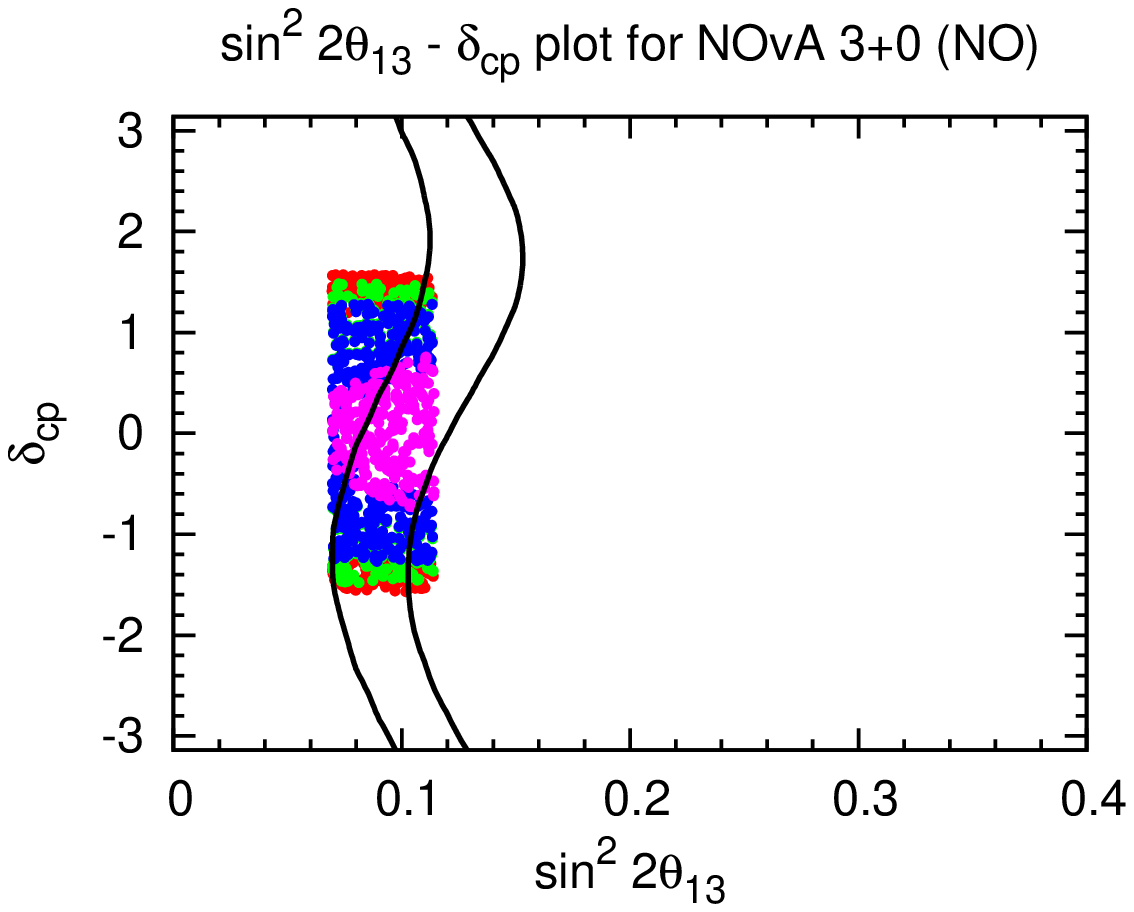}
\includegraphics[width=7cm,height=6cm]{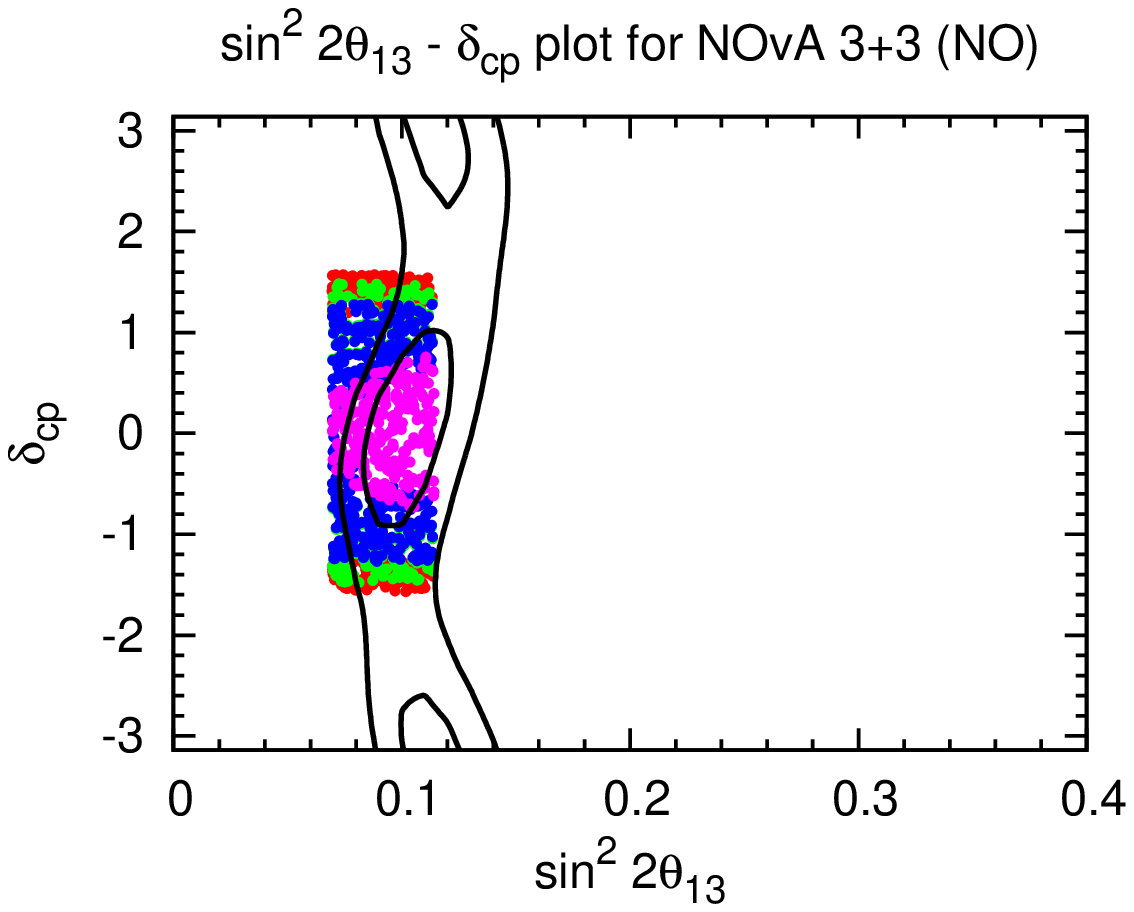}
\includegraphics[width=7cm,height=6cm]{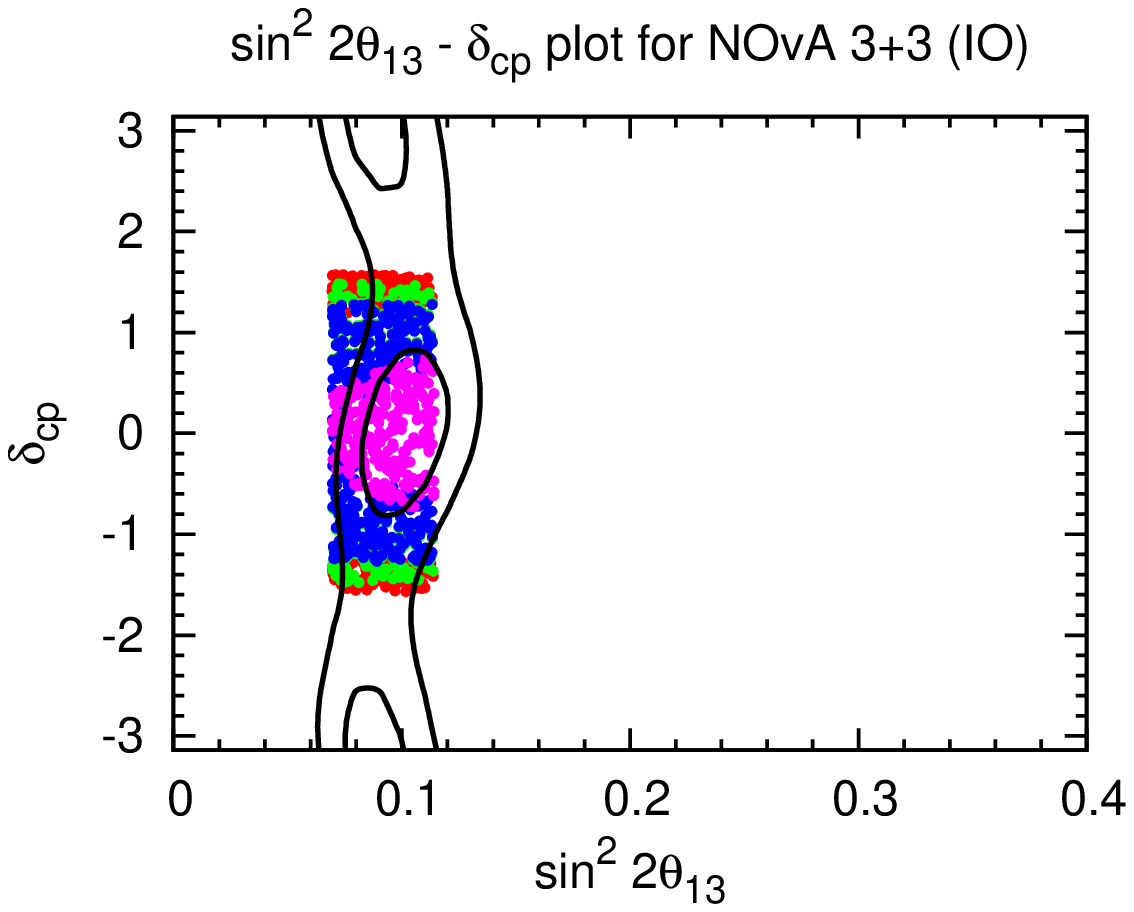}
\includegraphics[width=7cm,height=6cm]{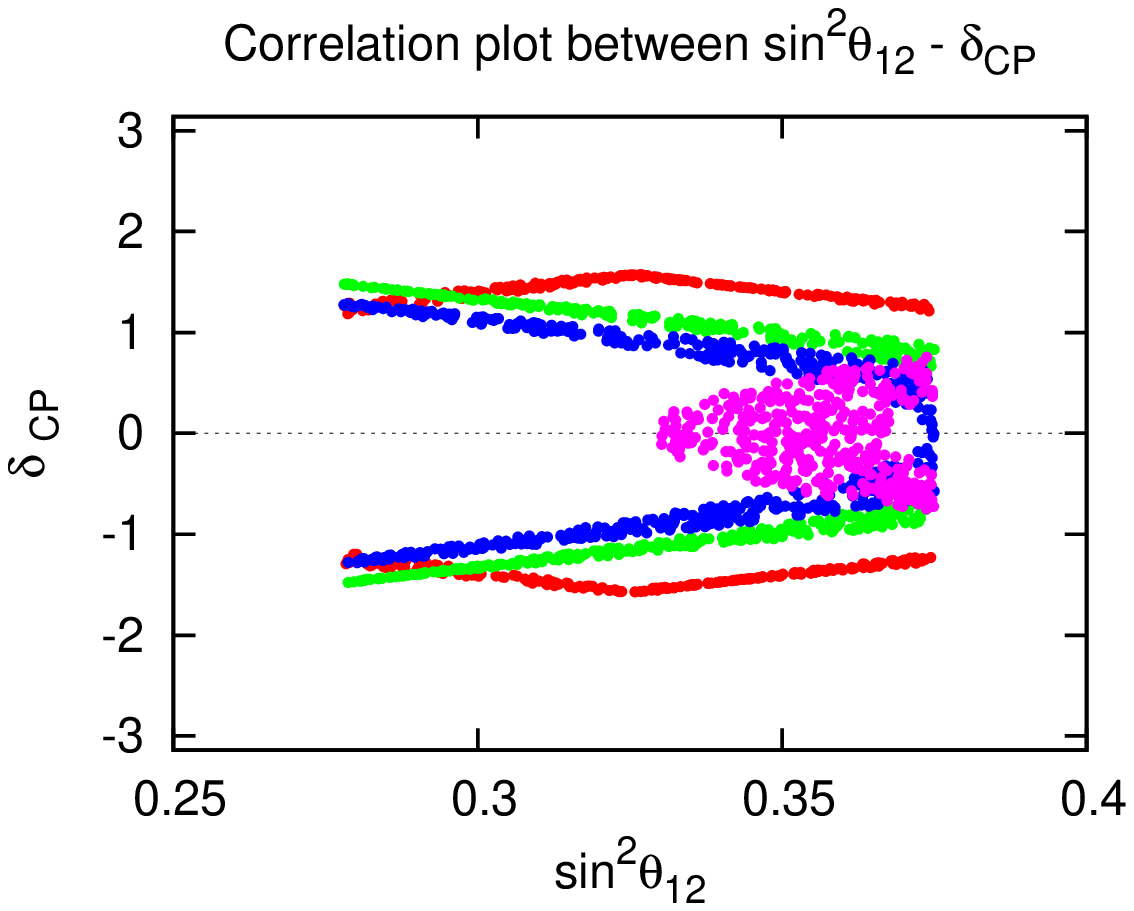}
\includegraphics[width=7cm,height=6cm]{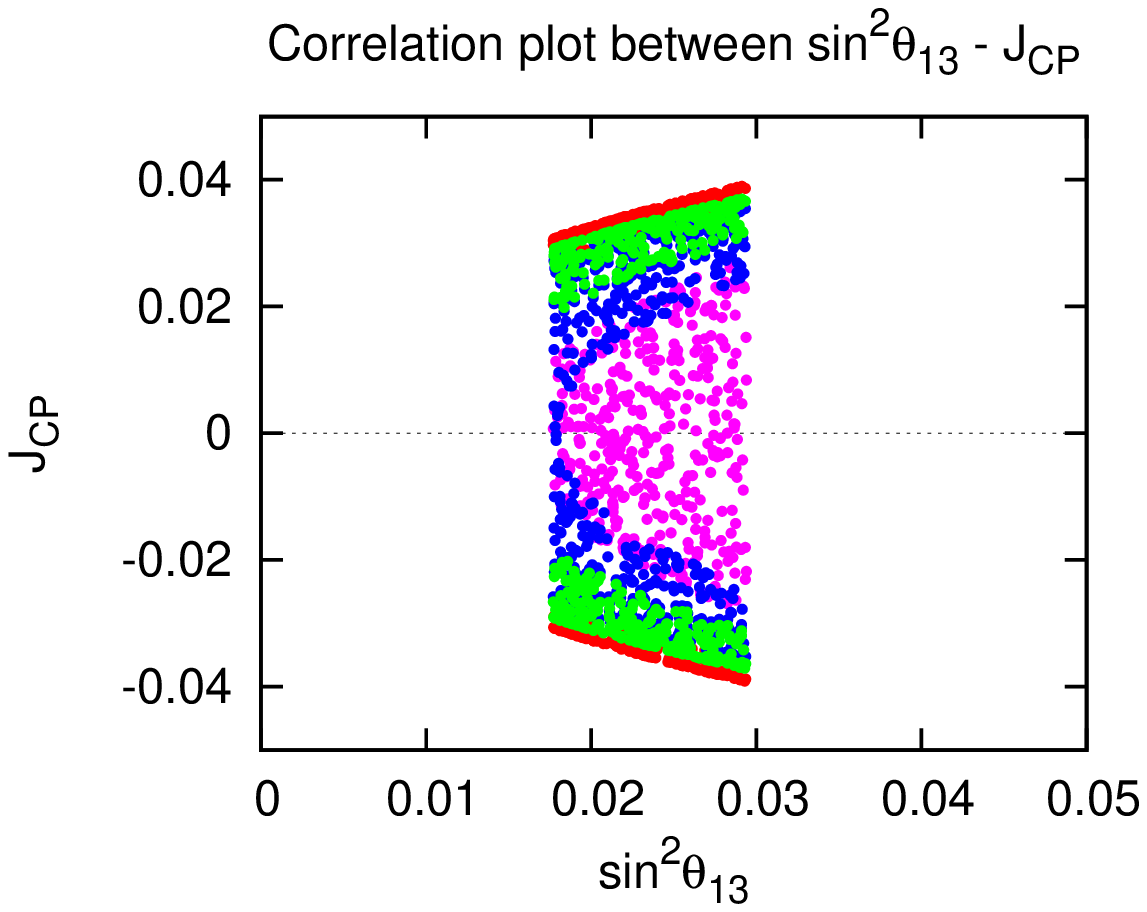}
\caption{Correlation plots between different observables due to 12 deviation in the charged lepton sector.
The top left panel represents the correlation plot between $\sin^2 \theta_{13}$ and $\sin^2\theta_{23}$.
The description of the other plots are indicated in the corresponding plot labels. In these plots the red, green, blue
and magenta regions 
correspond to TBM, GRA,  HG and  BM mixing patterns. The black solid lines
in the top (right panel) and the middle panel plots correspond to the experimentally allowed contours.}
\end{figure}
The Jarlskog invariant and the CP phase are found to be
\be
J_{CP}=\frac{1}{8 \sqrt 2}\Big (\sin 2 \theta_{12}^\nu \sin 2 \phi \sin \alpha \Big )\;,
\ee
\bea
\sin \delta_{CP}= \frac{\sin 2 \theta_{12}^\nu \sin \alpha (1+ \cos^2 \phi)}{2 \sqrt{Y}}\;.
\eea
Since the results for this deviation pattern are almost similar to the correction due to 12 rotation case, one obtains the same
constraints on $\delta_{CP}$ as in the previous case, which are listed in Table-2.

\section{Summary and Conclusion}

The recent observation of moderately large reactor mixing angle $\theta_{13}$  has ignited a lot of interest to understand the mixing
pattern in the lepton sector. It also  opens up promising perspectives for the observation of CP violation in the lepton sector. 
The precise determination  of $\theta_{13}$, in addition to providing a complete picture of neutrino mixing pattern
could be a signal of underlying physics responsible for lepton mixing and for the physics beyond the standard model.
In this context a number of neutrino mixing patterns like TBM/BM/GRA etc, were proposed  based on some discrete flavor symmetries like
$S_3$, $A_4$, $\mu-\tau$, etc. However, these symmetry forms of the mixing matrices predict vanishing reactor  and maximal
atmospheric mixing angles. To  accommodate the observed value of relatively large $\theta_{13}$, these mixing 
patterns should be modified by including appropriate perturbations.
In this work, we have considered
the simplest case of such perturbation which involves only minimal set of new independent parameters, i.e., one rotation angle and one phase, 
(which basically corresponds to  perturbation induced by a single rotation),
and  found that it is possible to explain the observed
neutrino oscillation data with such corrections. The predicted values of $\delta_{CP}$ are expected to be supported by the data from
currently running NO$\nu$A experiment with (3$\nu$ +3${\bar \nu}$) years of data taking. We have also shown that it is possible to predict the value of CP phase
with such corrections. We have also found that sizable leptonic CP violation characterized by the Jarlskog invariant $J_{CP}$, i.e., $|J_{CP}| \sim
10^{-2}$ could be possible in these scenarios.

{\bf Acknowledgments}
SM, SC and KND  would like to thank University Grants Commission for financial support.
The work of RM was partly supported by the Council of Scientific and Industrial Research,
Government of India through grant No. 03(1190)/11/EMR-II.

\end{document}